\documentclass[aps,pre,twocolumn,nofootinbib,reprint,showpacs,superscriptaddress]{revtex4-1}

\usepackage{amssymb}
\usepackage{graphicx}
\usepackage{amsmath}
\usepackage{braket}
\usepackage{hyperref}
\usepackage{bbold}
\usepackage{xcolor}
\usepackage{enumerate}

\def\bbra#1{\mathinner{\langle\!\langle{#1}|}}
\def\kket#1{\mathinner{|{#1}\rangle\!\rangle}}

\usepackage{environ}
\makeatletter
\NewEnviron{Lalign}{\tagsleft@true\begin{align}\BODY\end{align}}
\makeatother

\def\<{\langle}

\def\H{ {\cal H} }
\def\G{ {\cal G} }
\def\L{ {\cal L} }
\def\J{ {\cal J} }

\def\I{ \mathbb{1} }

\begin{document}

\title{Coarse-grained hidden entropy production in partially inaccessible \\quantum jump trajectories}

\author{Max F. Frenzel}
\affiliation{Department of Applied Physics, The University of Tokyo, 7-3-1 Hongo, Bunkyo-ku, Tokyo, Japan}
\affiliation{Controlled Quantum Dynamics Theory Group, Imperial College London, Prince Consort Road, London SW7 2BW, UK}
\author{Takahiro Sagawa}
\affiliation{Department of Applied Physics, The University of Tokyo, 7-3-1 Hongo, Bunkyo-ku, Tokyo, Japan}
\date{\today}

\begin{abstract}
We consider an open quantum system for which only a subset of all possible transitions are accessible, while the remaining ones are hidden from direct observation. Using a modification of the notion of quantum jump trajectories we introduce the coarse-grained hidden entropy, which quantifies the entropy production in the hidden subsystem conditioned on our observations of the visible part. The entropy production consisting of the sum of visible and coarse-grained hidden entropy is shown to satisfy an integral fluctuations theorem. Depending on the nature of the continuously occurring measurement and feedback processes between the interacting subsystems, the visible entropy can take negative values in which case the hidden systems acts as a Maxwell's demon. Our results allow us to study quantum-mechanical effects in autonomous systems, such as the autonomous quantum Maxwell's demon we introduce as an illustrative example.
\end{abstract}


\maketitle

\section{Introduction\label{sec:Introduction}}

Thermodynamics is one of the oldest and most well established areas of physics \cite{maxwell1872theory,Fermi1956}, and despite its long history it is still a very active field of research. In recent decades, one of the biggest breakthroughs in both classical and quantum non-equilibrium thermodynamics has been the discovery of fluctuation theorems (FTs), such as the celebrated Jarzynski \cite{Jarzynski1997} and Crooks \cite{Crooks1999_fluct} equalities. Fluctuation theorems are important mathematical tools linking very well understood (and idealized) equilibrium thermodynamics to the messier realities of non-equilibrium thermodynamics. They relate the probability of a process happening forwards in time with the probability of the reverse process, and form a generalization of one of the most fundamental laws in physics, the second law of thermodynamics \cite{Lieb1999}. Fluctuation theorems reveal the non-zero probability of the ``time-reversal of an irreversible process'', showing that entropy only increases on average, but can stochastically also decrease, especially in small systems. 

Most of the recent advances were made largely thanks to the advent of stochastic thermodynamics \cite{Seifert2012,Horowitz2014a,Hartich2014}. This framework broadens our understanding of thermodynamic concepts from the level of ensemble averages to individual microscopic trajectories in phase space. Even more recently, the extension of stochastic thermodynamics from the classical to the quantum regime, chiefly based on the concept of quantum jump trajectories \cite{Wiseman1996,Plenio1998a,Horowitz2012,Hekking2013,Strasberg2013,Horowitz2013,Elouard2015,Pigeon2016}, has lead to new insights into the thermodynamic behaviour of quantum systems. These insights have also contributed to results capturing truly quantum effects, such as the thermodynamic role of entanglement and coherences \cite{Renner2011,Aberg2011,Horodecki2011,Leggio2013,Brandao2013,Korzekwa,Lostaglio2014,Lostaglio2015,Anders,Elouard2016a,Jevtic2015}, some of which are even within current experimental reach \cite{Browne2016}.

Feedback control is another important idea which has recently received a great amount of interest, both in the classical \cite{Horowitz2010,Sagawa2011,Ashida2014,Sagawa2012a,Horowitz2011,Ito2013,Horowitz2014a} as well as the quantum regimes \cite{Funo2013,Horowitz2014,Jacobs2009,Sagawa2008,Gong2016}. Feedback control refers to performing measurements of a systems during a given protocol and then altering the remaining protocol based on the measurement outcome. In this way, it provides a useful tool to achieve the desired dynamics in small thermodynamic systems. 

Both fluctuation theorems and feedback control are well established in classical thermodynamics (although many open questions still remain), but much less is known about them in the quantum setting. One unsolved problem relates to fluctuation theorems for autonomous quantum systems \cite{Frenzel2014d,Frenzel2015,Malabarba2014a,Silva2016}. Currently, most quantum fluctuation theorems and feedback control protocols assume some kind of external control. Physically, this implies that there has to be a classical external system which is not part of the quantum description but which interacts with the quantum systems. Because this external system is not a part of the explicit quantum framework, it can neither develop correlations with the other systems, suffer from degradations, nor exhibit any other quantum effects. Such a treatment is justified if one assumes that this additional system is of macroscopic size, but this approach begins to fail as the auxiliary system itself gets smaller and quantum effects come into play. If we want to understand fully self-contained quantum machines which have a feedback controller embedded, an extension of current results is required.

This extension to autonomous quantum systems is the main aim of this article. Specifically we consider an open quantum system for which only a subset of its environmental interactions are experimentally accessible, similar to the classical setup discussed in \cite{Shiraishi2015b,Shiraishi2015}. Having split the system into visible and hidden parts, we find a process that can be thought of as continuous measurement and feedback performed by each system on the other. We introduce a coarse-grained hidden entropy which quantifies the entropy production in the hidden part based on the knowledge of the visible part, and establish a fluctuation theorem for the full system. We further show that the hidden system can act as an autonomous Maxwell's demon \cite{Horowitz2014a,Shiraishi2015,Camati2016,Bub2001,Maruyama2009,Chapman2015}, inducing a negative heat flow in the visible system in seeming violation of the second law, which can be rectified by including the coarse-grained hidden entropy in the account of entropy.

This article is structured as follows: In section \ref{sec:QJ} we review the basic idea of quantum jump trajectories and the FT introduced in \cite{Horowitz2013}. In section \ref{sec:ModifiedQJ} we introduce a modified version of the quantum jump approach which is suitable to partially hidden dynamics, introduce the coarse-grained hidden entropy $\Delta \sigma_Y$, and establish the general FT. These abstract results are then applied in section \ref{sec:Demon} to a simple but very informative toy model of two interacting qubits, one of which can act as a Maxwell's demon on the other. Finally we conclude with a discussion in section \ref{sec:Outlook}.

\section{Quantum Jump Trajectories\label{sec:QJ}}

The standard quantum jump (QJ) approach starts from one of the most universal description of the evolution of open quantum systems, the Lindblad master equation
\begin{equation}\label{eq:ME}
\dot{\rho} = -i[H,\rho] - \frac{1}{2} \sum_k \Bigl( L_k^{\dagger} L_k \rho + \rho L_k^{\dagger}L_k - 2L_k\rho L_k^{\dagger} \Bigr)
\end{equation}
where $\rho$ is the state of the quantum system, $H$ its Hamiltonian, and $L_k$ a set of Lindblad operators describing the dissipation due to interactions with the environment. Note that since our main interest is autonomous quantum systems, here and in the following we do not assume any time dependence in $H$ or the $L_k$, but all results easily generalize to time-dependent (i.e. non-autonomous) systems.

The master equation (\ref{eq:ME}) provides a description for the average evolution of a system coupled to an environment. We can go beyond this average description by choosing a particular \textit{unraveling} of the master equation. The notion of unraveling has recently gained very physical meaning with advances in experimental techniques that allow us to perform precise measurements at the level of individual quantum trajectories \cite{Kuhr2007,Barreiro2011,Murch2013,Roch2014,Weber2014}. The Lindblad operators $L_k$ are in general not unique, but choosing a particular set of measurements fixes them, and thus the possible jumps induced in the system. Specifically, we consider a setup in which we monitor all the system's energy exchanges with its environment, so that each $L_k$ corresponds to an excitation or relaxation in the system, a quantum jump. 

In the absence of any jump the system evolves according to the non-Hermitian effective Hamiltonian \mbox{$H_{{\rm eff}} := H - i\sum_k \frac{1}{2} L_k^{\dagger} L_k$} which acts as the generator of the non-unitary evolution operator \mbox{$U_{{\rm eff}}(t) = \exp(-i H_{{\rm eff}} t)$}. If the systems starts in the state $\rho(t)$ and we do not observe any jump for a time $\Delta t$, the system evolves to 
\begin{equation}\label{eq:evoNJ}
\rho_{{\rm nj}}(t+\Delta t)= \mathcal{U}_{\Delta t} (\rho(t)) := U_{{\rm eff}}(\Delta t) \rho(t) U_{{\rm eff}}^{\dagger}(\Delta t),
\end{equation} 
where the subscript ${\rm nj}$ stands for ``no jump''. This evolution contains both the free evolution due to $H$, as well as the correction to the free evolution due to our knowledge that no jump occurred. Due to its non-unitarity the map $\mathcal{U}_{\Delta t}$ is not trace-preserving, and the trace ${\rm tr}[\rho(t+\Delta t)]$ encodes the probability that no jump takes place during the interval $\Delta t$.

If on the other hand we do observe a jump in the short interval $dt$, corresponding to the operator $L_k$, the state of the system is discontinuously changed to
\begin{equation}
\rho_k(t+dt) = \mathcal{J}_{k} (\rho(t)) := L_k\rho(t) L_k^{\dagger} dt.
\end{equation}
As in the case of no jump, the trace of $\rho_k(t+dt)$ encodes the probability of the jump $k$ happening in the interval $dt$,
\begin{equation}
\delta p_k(t) = {\rm tr}[L_k^{\dagger} L_k\rho(t) ]dt.
\end{equation} 

Taking the ensemble average we get back to the original evolution predicted by the master equation, showing the consistency of the approach. However, beyond this the unraveling allows us to describe individual trajectories. Let us define the quantum jump trajectory $\tau := \{ \ket{a} ; (k_1,t_1) ;  ... ; (k_N,t_N) ; (\ket{b}, T) \}$ where at time $t=0$ we measure the system to be in state $\ket{a}$, then observe $N$ consecutive jumps labelled by $k_n$ occurring at time $t_n$, and finally at time $T$ perform another measurement where we find the system in $\ket{b}$. Defining the time intervals $\Delta t_n := t_{n+1} - t_{n}$, with $t_0 := 0$ and $t_{N+1} := T$, the probability of the particular trajectory $\tau$ conditioned on starting in state $\ket{a}$ is given by consecutive application of evolutions $\mathcal{U}_{\Delta t_n}$ and jumps $\mathcal{J}_{k_n}$ as
\begin{eqnarray} \label{eq:FullForwardProb}
P(\tau | a) &=& {\rm tr}\left[ \ket{b}\bra{b} \mathcal{U}_{\Delta t_N} \left(\mathcal{J}_{k_N}\left( ... \mathcal{J}_{k_1}\left( \mathcal{U}_{\Delta t_0}\left(\ket{a}\bra{a}\right)\right)\right)\right)\right] \nonumber \\
&=& \Bigl| \braket{b | U_{{\rm eff}}(\Delta t_N) ... L_{k_1}  U_{{\rm eff}}(\Delta t_0) | a} \Bigr|^2 dt^N.
\end{eqnarray}
In order to establish a fluctuation theorem we also have to define a backward trajectory $\tilde{\tau}$ corresponding to a time-reversal of the trajectory $\tau$. 

The Lindbland operators generally come in pairs, e.g. one corresponding to an excitation and one to a de-excitation. We denote the complementary operator of $L_k$ by $L_{\tilde{k}}$. As shown in \cite{Horowitz2013}, these pairs satisfy a kind of operator version of the local detailed balance condition,
\begin{equation} \label{eq:DetailedBalance}
L_k = L_{\tilde{k}}^{\dagger} e^{\Delta s_k / 2}
\end{equation} 
where $\Delta s_k = - \Delta s_{\tilde{k}}$ is the environmental entropy production associated with the jump $k$.

A jump $k$ in the forward trajectory $\tau$ at time $t$ has a conjugate jump $\tilde{k}$ at time $T-t$ in the backward trajectory $\tilde{\tau}$. Further, denoting the anti-unitary time-reversal operator by $\Theta$, we have for the time-reversal of (time-independent) operators $\tilde{O} := \Theta O \Theta^{-1}$ and states $\ket{\tilde{\psi}} := \Theta \ket{\psi}$. The reverse trajectory conjugate to $\tau$ thus starts in the time-reversed final state $\ket{\tilde{b}}$, then proceeds in reverse order through the conjugate jumps, and finally ends in the time-reversed initial state $\ket{\tilde{a}}$, i.e. $\tilde{\tau} = \{ \ket{\tilde{b}} ; (\tilde{k}_N,T-t_N) ;  ... ; (\tilde{k}_1,T-t_1) ; (\ket{\tilde{a}}, T) \}$. Following the same reasoning as for the forward trajectory, we find for the probability of the backward trajectory conditioned on starting in the state $\ket{\tilde{b}}$
\begin{eqnarray} \label{eq:FullBackProb}
P(\tilde{\tau} | \tilde{b}) &=& \Bigl| \braket{\tilde{a} | \tilde{U}_{{\rm eff}}(\Delta t_0) ... \tilde{L}_{\tilde{k}_N}  \tilde{U}_{{\rm eff}}(\Delta t_N) | \tilde{b}} \Bigr|^2 dt^N \nonumber \\
&=& \Bigl| \braket{b | U_{{\rm eff}}(\Delta t_N) ...  L_{\tilde{k}_1}^{\dagger}  U_{{\rm eff}}(\Delta t_0) | a} \Bigr|^2 dt^N \nonumber \\
&=& P(\tau | a) e^{-\Delta s_{{\rm env}}(\tau)}
\end{eqnarray}
where in the second line we have used \mbox{$\tilde{U}_{{\rm eff}}(\Delta t) = \Theta{U}_{{\rm eff}}^{\dagger}(\Delta t)\Theta^{-1}$} and then cancelled all the time-reversal operators appearing in pairs $\Theta \Theta^{-1}=\mathbb{1}$, and in the final line we have employed the detailed balance Eq. (\ref{eq:DetailedBalance}). The quantity $\Delta s_{{\rm env}}(\tau)$ is the environmental entropy production associated with the full forward trajectory and is given by $\Delta s_{{\rm env}}(\tau) := \sum_{n=1}^N \Delta s_{k_n}$.

Taking the ratio of Eqs. (\ref{eq:FullForwardProb}) and (\ref{eq:FullBackProb}) it is trivial to arrive at the detailed fluctuation theorem
\begin{equation} \label{eq:DetailedFT}
\frac{P(\tau | a)}{P(\tilde{\tau} | \tilde{b})} = e^{\Delta s_{{\rm env}}(\tau)}.
\end{equation} 
Finally, noting that the unconditional probability of the forward trajectory is $P(\tau) =P(\tau | a)P(a)$ and similarly for the backwards trajectory, we arrive after rearranging and summing over all $\tau$ at the integral fluctuation theorem (IFT)
\begin{equation} \label{eq:IFT}
\Braket{e^{-\Delta s_{{\rm tot}}(\tau)}} = 1
\end{equation} 
where the total entropy production \mbox{$\Delta s_{{\rm tot}}(\tau) = \Delta s_{{\rm env}}(\tau) +  \Delta s_{{\rm sys}}(\tau)$} is the sum of the environmental entropy production associated with the trajectory, as well as the stochastic entropy change $\Delta s_{{\rm sys}}(\tau) = \log{P(b)} - \log{P(a)}$.

In the following section we will modify the ideas presented here in order to accommodate for a situation where part of the full trajectory is hidden, with a subset of the transitions, associated with a subset of the operators $L_k$, being hidden from direct observation.

\section{Modified QJ Approach and Coarse-Grained Hidden Entropy\label{sec:ModifiedQJ}}
In many systems we are not interested in all the possible transitions, such as in Maxwell's demon type setups which we will discuss in section \ref{sec:Demon}. In other systems we are due to the systems' complexity experimentally restricted to only observing a certain subset of them \cite{Balzani2006,Serreli2007,Yildiz2008,Toyabe2012,Lan2012} which we shall call visible transitions, with the remainder being referred to as hidden transitions. In this section we address the question of how we can describe such systems within the quantum jump framework, and what we can say about the entropy production associated with the hidden transitions which are not directly accessible to us. 

In the following we will use $X$ to refer to anything visible, while $Y$ labels hidden properties. In line with this we split the Lindblad operators $L_k$ into two sets, a set of visible transitions $\{L_{k}\}_{k\in k_X}$ and a set of hidden transitions $\{L_{k}\}_{k\in k_Y}$, such that $\{L_{k}\}_{k\in k_X} \bigcup \{L_{k}\}_{k\in k_Y}$ contains all possible transitions. One assumption we have to make in order for the following results to hold is that the splitting is such that it does not separate complementary transitions, i.e. if $k$ is in $k_X$ ($k_Y$), then $\tilde{k}$ is also in $k_X$ ($k_Y$). If an excitation process is visible (hidden), the corresponding reverse process of de-excitation must also be visible (hidden). This assumption is satisfied in many realistic scenarios. If we for example monitor the photon exchange of a system with a bosonic bath, the complementary absorption and emission events are the visible transitions, whereas all other possible transitions of the system are hidden.

A visible trajectory is now described by $\tau_X := \{ \ket{a} ; (k_1,t_1) ;  ... ; (k_M,t_M) ; (\ket{b}, T) \}$ where all the $k_1, ... , k_M$ (with $M\leq N$) are in $k_X$. Note that we still assume the initial measurement revealing the state $\ket{a}$ and the final measurement of state $\ket{b}$ are performed on the full system, even if the splitting of the transitions into visible parts leads to a bipartite splitting in the states of the system, as is the case in the example in section \ref{sec:Demon} and many other systems of interest (see e.g. \cite{Horowitz2014a} and references therein). In general, each visible trajectory $\tau_X$ is compatible with multiple, in some cases infinitely many, hidden trajectories $\tau_Y$. However, the observed transitions in $\tau_X$ still allows us to infer something about the likelihood of the different hidden trajectories and thus to make inferences about the entropy production in the hidden system conditioned on $\tau_X$. If for example two consecutive visible transitions $k_i$ and $k_{i+1}$ both correspond to excitations in a two-level system, we know that in between these visible events some hidden transition(s) must have occurred that resulted in a de-excitation of the system. As noted by the authors in \cite{Elouard2016a}, the baths in the quantum jump setting play a double role, exchanging energy with the systems but in the process also extracting information.

Let us briefly restate some ideas of the previous section. In between transitions (visible or hidden), the system freely evolves (non-unitarily) according to \mbox{$\rho_{{\rm nj}}(t+dt) = \rho(t) - i(H_{{\rm eff}} \rho(t) - \rho(t) H_{{\rm eff}}^{\dagger})dt$}, as can be seen for example by expanding Eq. (\ref{eq:evoNJ}) to first order in $dt$, whereas a transition $k$ results in the discontinuous jump to $\rho_k(t+dt) = L_k \rho(t) L_k^{\dagger} dt$, with the probability of free evolution and each of the jumps being encoded in the traces of the respective states. Hence, if we do not know whether some of the transitions, namely the hidden subset labelled by $k_Y$, have occurred or not, the system evolution is to the best of our knowledge given by a probabilistic mixture of free evolution and hidden transitions, $\rho(t+dt) = \rho_{{\rm nj}}(t+dt) + \sum_{k\in k_Y} \rho_k(t+dt)$ (keeping in mind that the respective probabilities are already contained in the non-normalised states themselves). This allows us to introduce an effective master equation describing the evolution in between visible jumps
\begin{equation}\label{eq:ModifiedME}
\dot{\rho} = - i(H_{{\rm eff}} \rho - \rho H_{{\rm eff}}^{\dagger}) + \sum_{k\in k_Y} L_k \rho L_k^{\dagger}.
\end{equation}
This non-trace-preserving effective master equation is equivalent to the full master equation but without the visible jumps in the final sum. The first term describes the free evolution as well as the correction based on our knowledge that no jump has taken place, whereas the second term describes the effect due to the possibility of hidden transitions occurring. 

In the following we will make use of a key identity from linear algebra, $vec(\sum_i A_i \rho B_i) = \sum_i (B_i^{T} \otimes A_i) vec(\rho)$ for operators $A_i$, $B_i$, and $\rho$, where $vec(\cdot)$ refers to the vectorisation of an ($n\times m$) matrix, i.e. stacking its columns on top of each other resulting in a vector of length $nm$. This allows us to move from Hilbert space to Liouville space where super-operators are represented by matrices acting on density operators represented by vectors. To simplify notation, we will denote the vectorised version of a density matrix $\rho$ by $\kket{\rho} := vec(\rho)$. Using the identity above we can transform the effective master equation (\ref{eq:ModifiedME}) into the compact Liouville space form
\begin{equation}\label{eq:ModifiedMELiouville}
\kket{\dot{\rho}} = \G \kket{\rho}
\end{equation}
where the superopetor matrix $\G$ is given by \mbox{$\G := \H + \L_{{\rm nj}} + \L_Y$}, with \mbox{$\H := -i(\mathbb{1}\otimes H - H^{T}\otimes\mathbb{1})$} corresponding to free evolution under the Hamiltonian $H$, \mbox{$\L_{{\rm nj}} := -\frac{1}{2}\sum_k(\mathbb{1}\otimes L_k^{\dagger}L_k+L_k^{T}L_k^*\otimes\mathbb{1})$} accounting for the correction due to the no-jump knowledge, and \mbox{$\L_Y := \sum_{k_Y} L_{k_Y}^* \otimes L_{k_Y}$} representing the hidden transitions. Eq. (\ref{eq:ModifiedMELiouville}) can be solved to give the state of the system after an evolution of time $\Delta t$ without visible transition as \mbox{$\kket{\rho(t+\Delta t)} = e^{\G \Delta t} \kket{\rho(t)}$}. Vectorising the action of a visible jump $k$ in the interval $dt$ we have the discontinuous change in the state vector \mbox{$\kket{\rho_k(t+dt)} = (L_{k}^* \otimes L_{k}) \kket{\rho(t)} dt := \J_k \kket{\rho(t)} dt$}.

\begin{figure}[tb]
\includegraphics[width=\columnwidth]{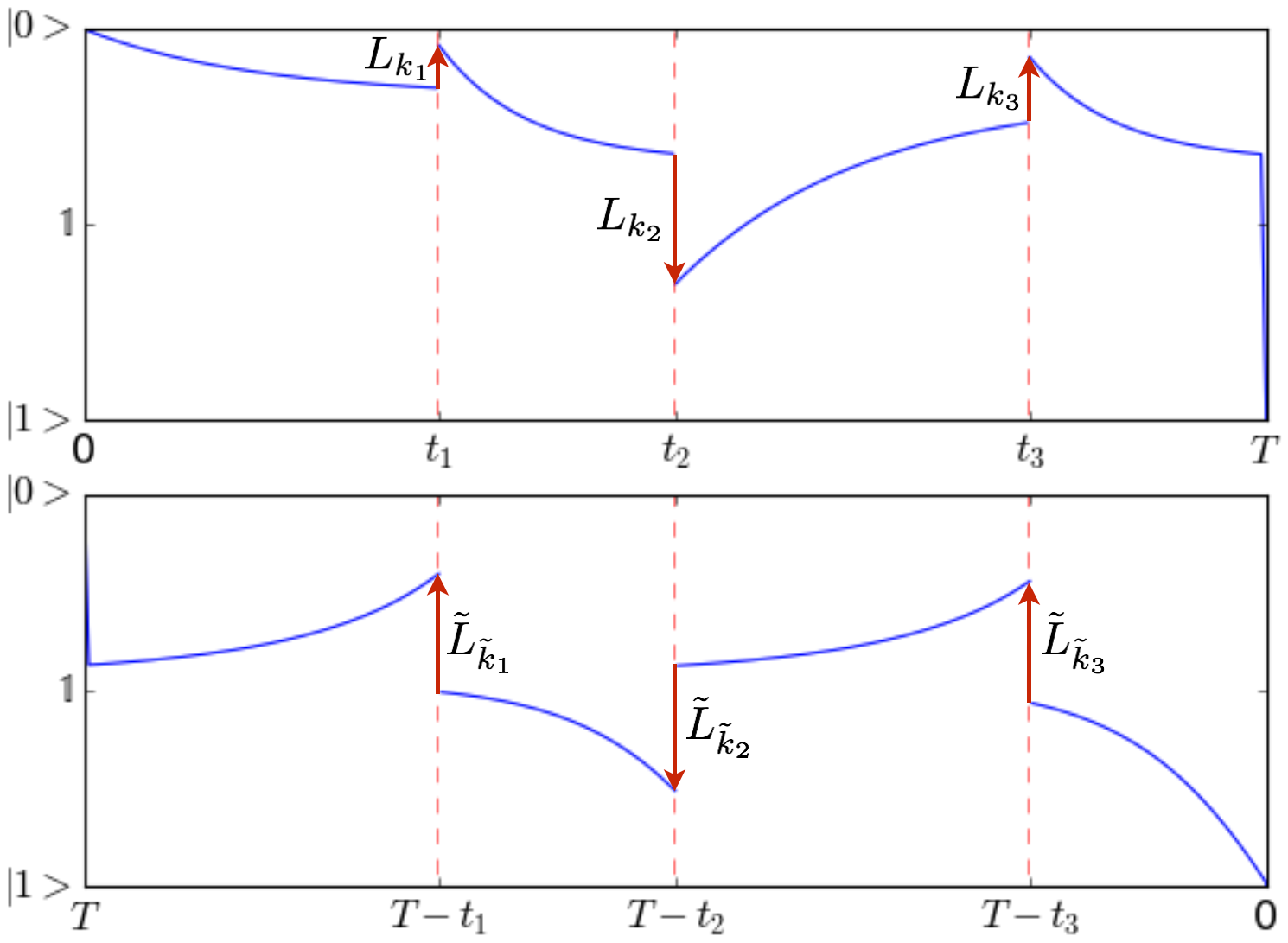}
\caption{\textbf{Hidden System Evolution.} Example evolution of a hidden system $Y$ conditioned on the visible forward trajectory $\tau_X$ (top) and corresponding backward trajectory $\tilde{\tau}_X$ (bottom, note the reversal of the time axis). Observations of visible jumps lead to discontinuous back-actions in the hidden system. Note that in this modified QJ approach, the hidden system does in general not stay in a pure state as it would for a standard QJ trajectory. This specific example corresponds to the visible trajectory $\tau_X = \{\ket{g,0}; (k_1 = 4, t_1 = 0.9); (k_2 = 1, t_2 = 1.5); (k_3 = 4, t_3 = 2.4); (\ket{e,1},T=3)\}$ with $\gamma_X = 0.5 = \gamma_Y$ and $\lambda = 0$ for the model introduced in section \ref{sec:Demon}. In this example $Y$ does not develop any coherences and its state conditioned on the knowledge of the visible trajectory is fully specified by the mixture of $\ket{0}$ and $\ket{1}$ shown above.\label{fig:trajectory}}
\end{figure}

Considering the visible trajectory $\tau_X$ we can now find the (non-normalised) state just before the final measurement at time $T$ by starting from state $\kket{a}$ (where we use the notation $\kket{a}:= vec(\ket{a}\bra{a})$ for brevity) and then consecutively applying evolution operators $e^{\G \Delta t_i}$ followed by visible jumps $\J_{k_{i+1}}$. Specifically, we have $\kket{\rho(T)} = e^{\G \Delta t_M}\J_{k_{M}} ... \J_{k_{1}}e^{\G \Delta t_0}\kket{a}dt^M$. The top half of Fig. \ref{fig:trajectory} shows an example evolution of the hidden part of the toy system we introduce in section \ref{sec:Demon}, where visible and hidden transitions induce a bipartite splitting in the system. This hidden part is found by converting $\kket{\rho(t)}$ back to the density matrix $\rho(t)$ and then tracing over the visible subsystem $\rho_Y(t) = {\rm tr}_X[\rho(t)]$. Note that while in the standard QJ approach pure states remain pure, this modified QJ approach leads in general to mixed states (or mixed states of sub-systems as in Fig. \ref{fig:trajectory} where the visible system remains pure throughout) due to the lack of knowledge of the hidden transitions.

Finally applying the measurement at time $T$ which results in the outcome $\kket{b}$, we arrive at the probability of the visible trajectory $\tau_X$ conditioned on starting in $\ket{a}$
\begin{equation}\label{eq:PVisibleForward}
P(\tau_X | a) = \bbra{b}  e^{\mathcal{G}\Delta t_{M}} \mathcal{J}_{k_{M}}...\mathcal{J}_{k_1} e^{\mathcal{G}\Delta t_0}\kket{a} dt^{M}.
\end{equation}
Note that if all transitions are visible and none are hidden, Eq. (\ref{eq:PVisibleForward}) reduces to Eq. (\ref{eq:FullForwardProb}) with $\tau_X = \tau$ as would be expected, showing the consistency of the approach.

Analogous to the considerations in the previous section, we define the complementary visible backwards trajectory $\tilde{\tau}_X := \{ \ket{\tilde{b}} ; (\tilde{k}_M,T-t_M) ;  ... ; (\tilde{k}_1,T-t_1) ; (\ket{\tilde{a}}, T) \}$. Considering the time reversal of all the operators involved, we see that in the reverse process \mbox{$\overline{\G}:= -\H + \L_{\rm nj} + \L_Y$}, or more specifically \mbox{$\overline{\G}^{\dagger} = \H + \L_{\rm nj} + \L_Y^{\dagger}$} play a similar role as $\G$ does for the forward process. Specifically, the probability of the visible backwards trajectory evaluates as
\begin{equation}\label{eq:PVisibleBackward}
P(\tilde{\tau}_X | \tilde{b}) = \bbra{b}  e^{\mathcal{\bar{G}^{\dagger}}\Delta t_{M}} \mathcal{J}_{k_{M}}...\mathcal{J}_{k_1} e^{\mathcal{\bar{G}^{\dagger}}\Delta t_0}\kket{a} dt^{M} e^{-\Delta s_{{\rm env}}(\tau_X)}
\end{equation}
with $\Delta s_{{\rm env}}(\tau_X) = \sum_{n=1}^M \Delta s_{k_n}$ being the environmental entropy production associated with the visible transitions. Detailed derivations can be found in Appendix \ref{app:TimeReversal}, but here simply note that unless all transitions are visible $\bar{\G}^{\dagger} \neq \G$. Thus taking the ratio of the forward and backward probability, the operator expressions do not simply cancel and leave us with $e^{-\Delta s_{{\rm env}}(\tau_X)}$ as in Eq. (\ref{eq:DetailedFT}) (unless all transitions are visible, then we recover the previous result). Instead we arrive at a modified detailed fluctuation theorem
\begin{equation} \label{eq:DetailedFTModified}
\frac{P(\tau_X | a)}{P(\tilde{\tau}_X | \tilde{b})} = e^{\Delta s_{{\rm env}}(\tau_X) + {\Delta \sigma}_Y (\tau_X)}
\end{equation} 
where we have introduced the new quantity 
\begin{equation} \label{eq:DeltaIY}
{\Delta \sigma}_Y (\tau_X) := \log\frac{ \bbra{b}  e^{\mathcal{G}\Delta t_{M}} \mathcal{J}_{k_{M}}...\mathcal{J}_{k_1} e^{\mathcal{G}\Delta t_0}\kket{a}}{\bbra{b}  e^{\mathcal{\bar{G}^{\dagger}}\Delta t_{M}} \mathcal{J}_{k_{M}}...\mathcal{J}_{k_1} e^{\mathcal{\bar{G}^{\dagger}}\Delta t_0}\kket{a}}
\end{equation} 
which we call the \textit{coarse-grained hidden entropy production} of $Y$ given the visible trajectory $\tau_X$. ${\Delta \sigma}_Y (\tau_X)$ quantifies the entropy production in the hidden part of the system based on the limited inference we can make from our observation of the visible part, and can take both positive and negative values. The ratio appearing in ${\Delta \sigma}_Y (\tau_X)$ can loosely be seen as the ratio of the densities of hidden trajectories $\tau_Y$ given information provided by the visible forward trajectory $\tau_X$ and visible backward trajectory $\tilde{\tau}_X$ respectively. As we will show in section \ref{sec:Demon}, if $\tau_X$ provides no information about the hidden system, ${\Delta \sigma}_Y (\tau_X)$ vanishes, whereas in cases where the visible trajectory allows for only one particular hidden trajectory $\tau_Y$, it reduces to the environmental entropy production $\Delta s_{{\rm env}}(\tau_Y)$ of that specific trajectory. In all intermediate situations it provides a kind of coarse-graining over the possible environmental entropy productions, conditioned on our limited knowledge. It is generally non-zero if $\tau_X$ reveals some forward-backward asymmetry in the compatible hidden trajectories.

In certain cases we can show explicitly that
\begin{eqnarray} \label{eq:CoarseGrained}
{\Delta \sigma}_Y (\tau_X) &=& - \log\sum_{\tau_Y} P(\tau_Y | \tau_X) e^{-\Delta s_{{\rm env}}(\tau_Y)}\nonumber\\
&:=& - \log\Braket{e^{-\Delta s_{{\rm env}}(\tau_Y)}}_{\tau_Y | \tau_X}
\end{eqnarray} 
which clearly shows the coarse-graining. We believe that Eq. (\ref{eq:CoarseGrained}) generally holds if the correct probabilities $P(\tau_Y | \tau_X)$ can be determined, but have only shown this explicitly for special cases (see section \ref{sec:Demon}). 

Note that the entropy production $\Delta \sigma(\tau_X) = \Delta s_{{\rm env}}(\tau_X) + {\Delta \sigma}_Y (\tau_X) + \Delta s_{{\rm sys}}(\tau_X)$, where we have used $\sigma$ instead of $s$ to explicitly show that $\Delta \sigma(\tau_X)$ contains a coarse-grained quantity, can be seen as an extension of the classical coarse-grained entropy productions presented in \cite{Kawai2007,Sagawa2012a}.

From Eq. (\ref{eq:DetailedFTModified}) it straightforwardly follows that the entropy production $\Delta \sigma$ again satisfies an IFT
\begin{equation} \label{eq:IFTModified}
\Braket{e^{-\Delta \sigma(\tau_X)}} = 1.
\end{equation} 
In the steady state, the average change in stochastic entropy vanishes, $\braket{\Delta s_{{\rm sys}}(\tau_X)} = 0$, and using Jensen's inequality and the convexity of the exponential we obtain the second-law like inequality
\begin{equation} \label{eq:SecondLaw}
\Braket{\Delta s_{{\rm env}}(\tau_X)} + \Braket{{\Delta \sigma}_Y (\tau_X)} \ge 0.
\end{equation} 
From this we see how the hidden transitions in $Y$ can act as an autonomous Maxwell's demon if $\braket{{\Delta \sigma}_Y (\tau_X)}>0$, feeding information into $X$, allowing it to seemingly violate the conventional second law. If on the other hand $\braket{\Delta s_{{\rm env}}(\tau_X)}>0$, the average hidden entropy often takes on a negative sign (see Figs. \ref{fig:entropy} and \ref{fig:SecondLaw}). Thus Eq. (\ref{eq:SecondLaw}) in many cases gives a tighter bound than the conventional second law.

In the following section we will apply the general results to a simple but very illustrative model of a two qubit system.

\section{Example: Two qubit Maxwell's Demon\label{sec:Demon}}
\begin{figure}[tb]
\includegraphics[width=\columnwidth]{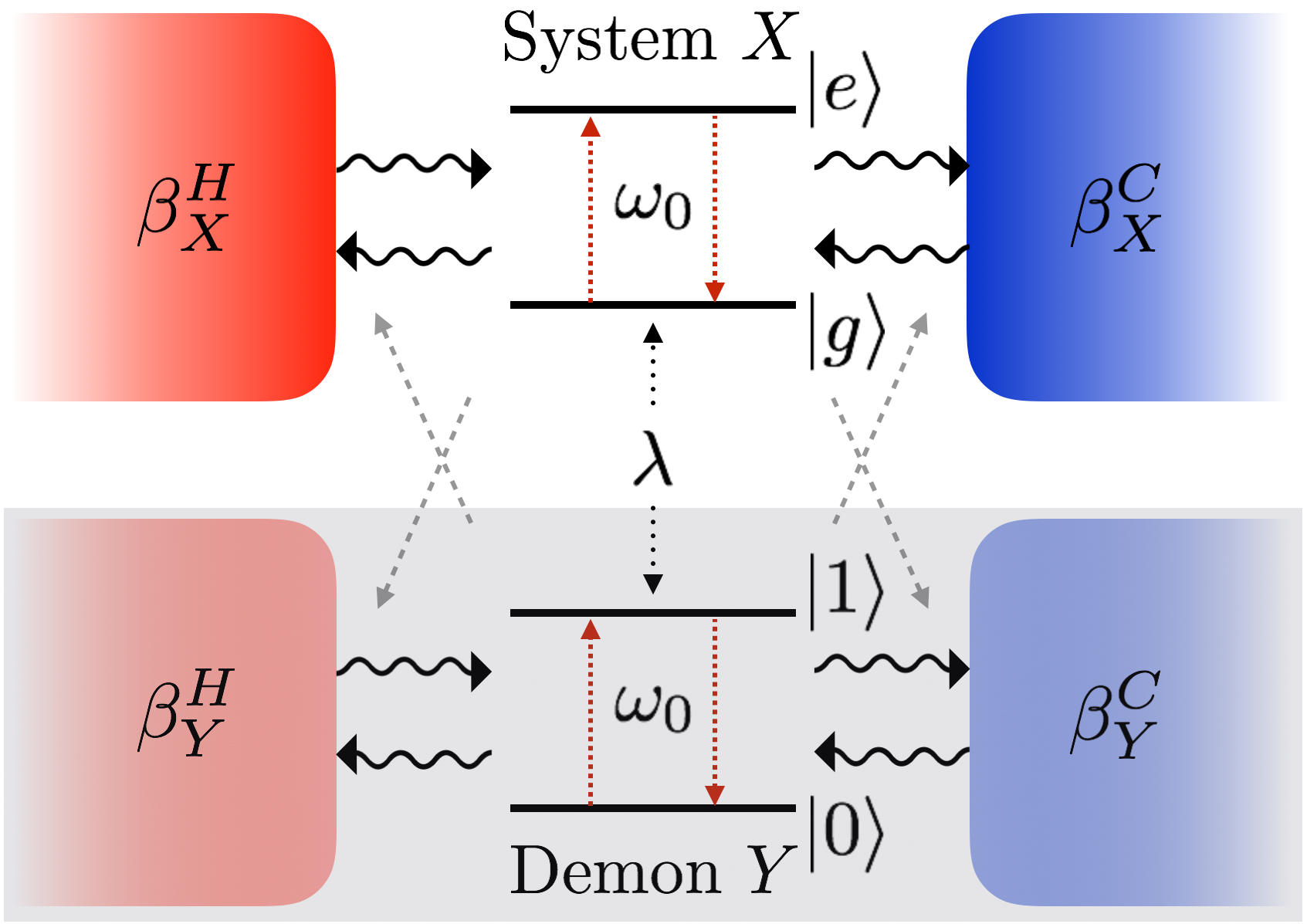}
\caption{\textbf{Two Qubit Autonomous Maxwell's Demon.} A system qubit $X$ and a demon qubit $Y$, each coupled to a hot and a cold reservoir, interact in such a way that the state of one system influences the reservoir coupling of the other system. Only the energy exchanges of $X$ with its baths can be monitored, while the demon $Y$ and its bath interactions are assumed to be not directly accessible to the experimenter.\label{fig:model}}
\end{figure}
Bipartite systems comprising two separate but interacting sub-systems are ubiquitous and of great interest, with numerous biological and artificial microscopic systems exhibiting a bipartite structure \cite{Tian2009,Ito2013,Hartich2014,Horowitz2014,Horowitz2014a,Kutvonen2015}. A particularly interesting class of bipartite systems are autonomous Maxwell's demons, where one subsystem measures and performs feedback on the other subsystem allowing it to seemingly violate the second law, for example by inducing a heat flow from a cold to a hot bath. 

Here we will illustrate our general results for the coarse-grained hidden entropy production using one of the simplest bipartite systems conceivable, two coupled two level systems. The basic setup, consisting of a system qubit $X$ with states labelled by $\ket{g}$ and $\ket{e}$, and a demon qubit $Y$ with states $\ket{0}$ and $\ket{1}$, is illustrated in Fig. \ref{fig:model}. The energy splitting of the two systems is assumed to be equal, given by the frequency $\omega_0$. The joint system living in the Hilbert space $\mathcal{H} = \mathcal{H}_X \otimes \mathcal{H}_Y$ has four states $\{\ket{g,0},\ket{g,1},\ket{e,0},\ket{e,1}\}$, similar to the classical autonomous Maxwell's demons discussed in \cite{Horowitz2014a,Shiraishi2015b}. The free Hamiltonian of the two qubits is given by \mbox{$H_{\rm free} = \frac{\omega_0}{2} (\sigma_z \otimes \mathbb{1} + \mathbb{1} \otimes \sigma_z)$}. We further assume that the systems may interact via \mbox{$H_{\rm int} = \lambda \bigl(\ket{g,1}\bra{e,0} + \ket{e,0}\bra{g,1}\bigr)$}, which can generate correlations between the two sub-systems. Since $[H_{\rm int},H_{\rm free}]=0$ the interaction is strictly energy conserving and we can unambiguously define local thermal states with respect to each system's free Hamiltonian.

\begin{figure}[tb]
\includegraphics[width=0.8\columnwidth]{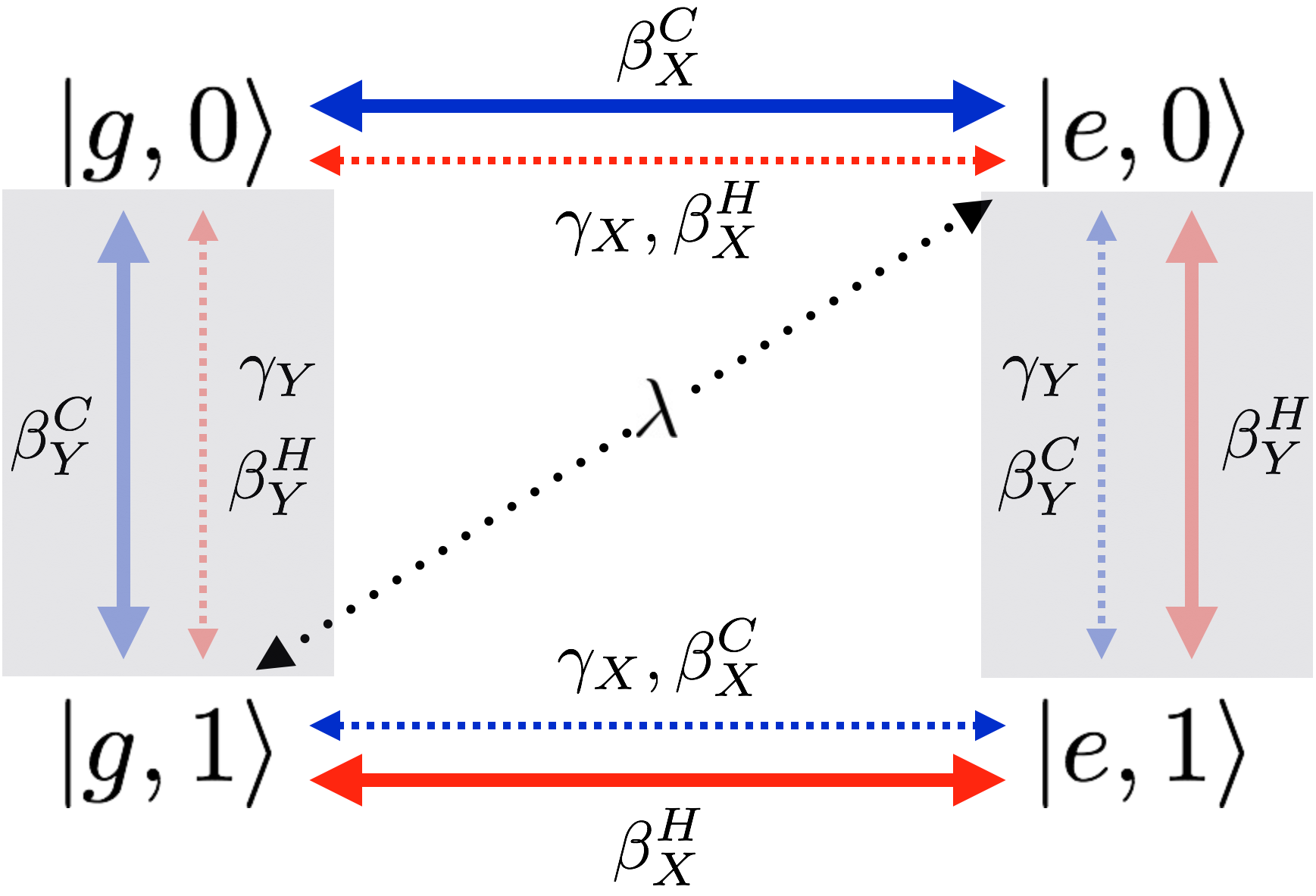}
\caption{\textbf{State and transition diagram.} The model in Fig. \ref{fig:model} leads to a bipartite transition diagram with four possible joint states. The horizontal system transitions are visible, while the vertical demon transitions are hidden. The (partial) suppression of certain transitions in one system conditioned on the state of the other system (captured by the factors $\gamma_X, \gamma_Y\in [0,1]$) leads to a continuous measurement and feedback process between the systems. The degenerate states $\ket{e,0}$ and $\ket{g,1}$ can additionally be connected via an interaction term in the Hamiltonian with coupling strength $\lambda$.\label{fig:state}}
\end{figure}

Both qubits $X$ and $Y$ are coupled to two thermal baths, each to a bath at hot (inverse) temperatures $\beta_X^H$ and $\beta_Y^H$, and to a bath at cold temperatures $\beta_X^C$ and $\beta_Y^C$ respectively. Crucially, the coupling between the qubits and their respective baths is not a simple direct coupling, but is influenced by the state of the other qubit, thus allowing one system to act as a Maxwell's demon for the other by performing autonomous and continuous measurement and feedback. If the demon is in state $\ket{0}$ ($\ket{1}$) we want the system to preferentially interact with its cold (hot) bath, and similarly if the system is in state $\ket{g}$ ($\ket{e}$) we want the demon to preferentially  interact with its cold (hot) bath, with the undesired interactions suppressed by factors $\gamma_X, \gamma_Y \in [0,1]$ respectively. The resulting state and transition diagram is shown in Fig. \ref{fig:state}.

This conditioned bath coupling leads to a continuous information flow between the systems. The factors $\gamma_X$ and $\gamma_Y$ allow us to tune the strength of feedback, with $\gamma_X, \gamma_Y = 0$ being perfect feedback, and $\gamma_X, \gamma_Y = 1$ resulting in completely independent systems. As we increase $\gamma_X$ ($\gamma_Y$) from zero to unity, the system (demon) receives increasingly less information from the demon (system). Simultaneously, the demon's (system's) feedback on the system (demon) weakens. This illustrates the continuous process of mutual measurement and feedback and the resulting information flow in the autonomous system. Looking at Fig. \ref{fig:state} one can readily convince oneself that if for example $\beta_Y^C > \beta_X^C$ and $\beta_Y^H < \beta_X^H$ (greater temperature difference in the demon than the system) and $\gamma_X=0=\gamma_Y$ (perfect feedback), the system will on average transition clockwise through the state diagram and we observe a negative heat flow from the system's cold bath to its hot bath, which is compensated by the positive heat flow through the demon.

The full system evolution including couplings to the four baths can be described by the Lindblad master equation (\ref{eq:ME}) with jump operators
\begin{eqnarray}\label{eq:LSystem}
L_1 = \gamma_1 \sigma_- \otimes (\ket{1}\bra{1} + \gamma_X\ket{0}\bra{0}) \nonumber\\
L_2 = \gamma_2 \sigma_+ \otimes (\ket{1}\bra{1} + \gamma_X\ket{0}\bra{0})\nonumber\\[0.25cm]
L_3 = \gamma_3 \sigma_- \otimes (\gamma_X\ket{1}\bra{1} + \ket{0}\bra{0}) \nonumber\\ 
 L_4 = \gamma_4 \sigma_+ \otimes (\gamma_X\ket{1}\bra{1} + \ket{0}\bra{0})
 \end{eqnarray}
 and
 \begin{eqnarray}\label{eq:LDemon}
L_5 = \gamma_5(\ket{e}\bra{e}+\gamma_Y\ket{g}\bra{g}) \otimes \sigma_- \nonumber\\ 
L_6 = \gamma_6(\ket{e}\bra{e}+\gamma_Y\ket{g}\bra{g}) \otimes \sigma_+\nonumber\\[0.25cm]
L_7 = \gamma_7(\gamma_Y\ket{e}\bra{e}+\ket{g}\bra{g}) \otimes \sigma_- \nonumber\\ 
L_8 = \gamma_8(\gamma_Y\ket{e}\bra{e}+\ket{g}\bra{g}) \otimes \sigma_+
\end{eqnarray}
where $\sigma_{+} = \ket{e}\bra{g}$ and $\sigma_{-} = \ket{g}\bra{e}$ for the system, and similarly for the demon. As noted above, these operators come in pairs, odd subscripts labelling relaxations and even numbers excitations. The label pair $(1,2)$ corresponds to the system's hot bath, $(3,4)$ its cold bath, and similarly with $(5,6)$ and $(7,8)$ for the demon. The coupling strengths $\gamma_i$ are given by $\gamma_1 = \sqrt{\Gamma_{X}^{H} (\bar{n}_{X}^{H}+1)}$, $\gamma_2 = \sqrt{\Gamma_{X}^{H} \bar{n}_{X}^{H}}$, and similar expressions for the other pairs. The factors $\Gamma_{i}^{j}$ are the actual coupling strength to every bath (which we will in the remainder all set to unity for simplicity) and $\bar{n}_{i}^{j} = (e^{\beta_i^j \omega_0}-1)^{-1}$ is the mean occupation number of bath $j\in(H,C)$ of system $i\in(X,Y)$. One can readily confirm that these Lindblad operators satisfy the operator detailed balance Eq. (\ref{eq:DetailedBalance}) with environmental entropy productions 
\begin{eqnarray}\label{eq:EnvEntX}
\Delta s_1 = \beta_X^H \omega_0 = -\Delta s_2 \nonumber\\
\Delta s_3 = \beta_X^C \omega_0 = -\Delta s_4 
 \end{eqnarray}
for the system, and
 \begin{eqnarray}\label{eq:EnvEntY}
\Delta s_5 = \beta_Y^H \omega_0 = -\Delta s_6 \nonumber\\
\Delta s_7 = \beta_Y^C \omega_0 = -\Delta s_8
\end{eqnarray}
for the demon. The Lindblad master equation (\ref{eq:ME}) with jump operators (\ref{eq:LSystem}) and (\ref{eq:LDemon}) can be directly derived from a microscopic description of the four heat baths. In the absence of the coupling term proportional to $\lambda$ the Born-Markov approximation on which the master equation is based holds unambiguously. For interacting composite systems, i.e. if $\lambda \neq 0$, the question whether this approximation is still accurate becomes more complicated (see e.g. \cite{Nakatani2010} for a discussion of this issue), but for our purpose here, particularly for small $\lambda$, it does provide a sufficient approximation.

To get to the main aim of this article, the study of hidden entropy production, and to apply our general results, we now split the jump operators into a visible and a hidden set. The system transitions labelled by $k_X = \{1,2,3,4\}$ are assumed to be experimentally accessible, whereas the demon transitions $k_Y = \{5,6,7,8\}$ are inaccessible, and can at best only be (partially) inferred from observations of the visible transitions. Note that in this case the splitting into visible and hidden transitions also coincides with a splitting into visible and hidden systems. This bipartite splitting is satisfied in many real systems of interest, but is not necessary for the main results to hold. They are equally applicable if we for example assumed that only one of the four baths was hidden.

We can now consider the coarse-grained hidden entropy production in the demon ${\Delta \sigma}_Y$ Eq. (\ref{eq:DeltaIY}) and the IFT Eq. (\ref{eq:IFTModified}). We find three distinct regimes depending on the type of feedback, controlled by the parameters $\gamma_X$ and $\gamma_Y$ which we will discuss individually in the following.

\subsection{Fully Reconstructible Hidden Trajectory}
\begin{figure}
\includegraphics[width=\columnwidth]{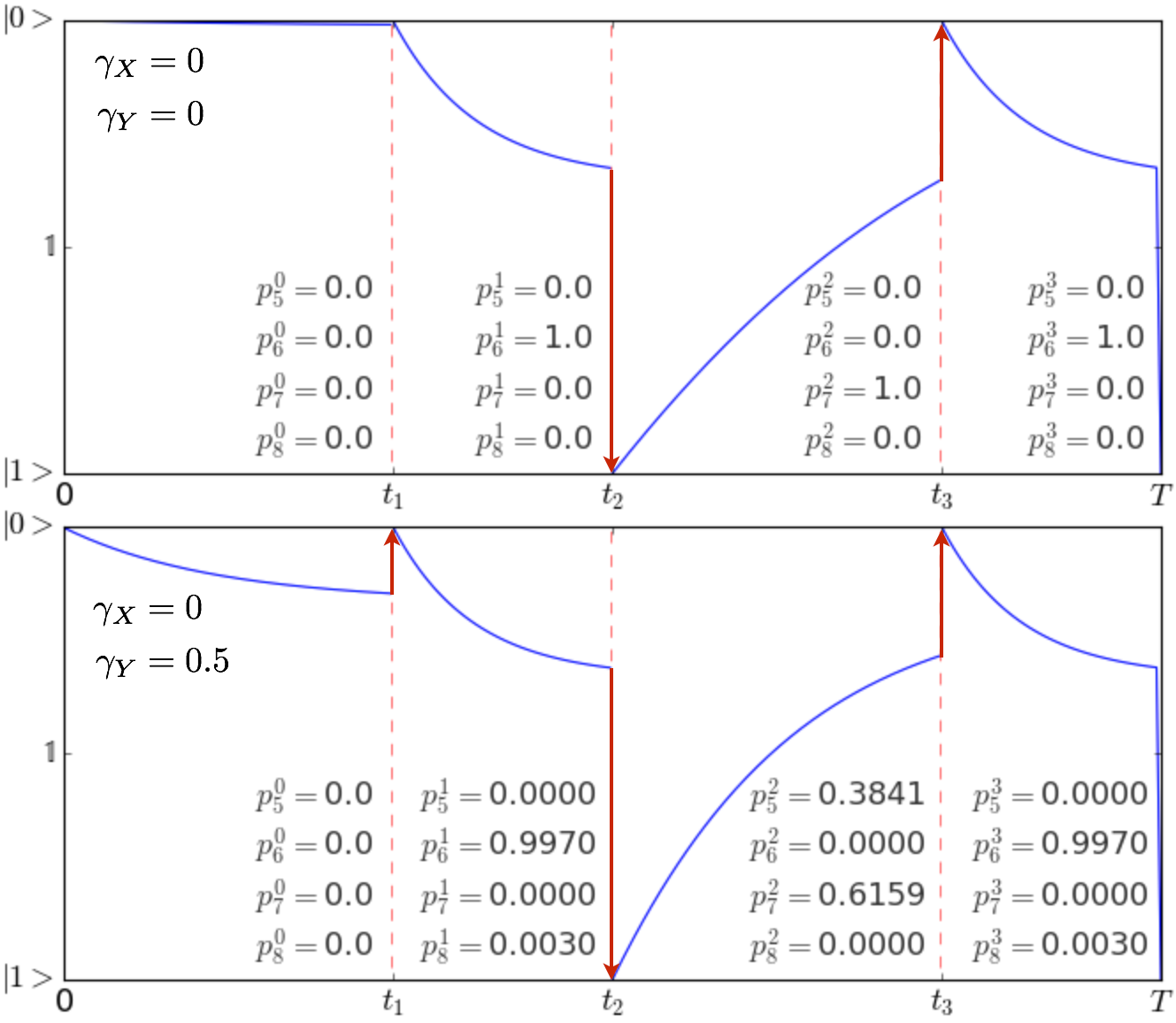}
\caption{\textbf{Hidden evolution in (partially) reconstructible scenario.} Evolution of the hidden demon $Y$ conditioned on the observed trajectory $\tau_X = \{\ket{g,0}; (k_1 = 4, t_1 = 0.9); (k_2 = 1, t_2 = 1.5); (k_3 = 4, t_3 = 2.4); (\ket{e,1},T=3)\}$ with $\lambda = 0$. The top half shows the case where both $\gamma_X$ and $\gamma_Y$ are zero, corresponding to perfect feedback. In the lower half $\gamma_Y=0.5$ implying reduced feedback from the system to the demon (or equivalently reduced measurement of the system by the demon). Note that in both cases the backaction from the visible transition brings the demon back to a pure state (this is no longer true if $\gamma_X>0$, c.f. Fig. \ref{fig:trajectory} and Fig. \ref{fig:bloch}). For $\gamma_Y=0$ (top) the knowledge of $\tau_X$ allows us to determine exactly which, if any, hidden transition must have occurred between visible jumps, and thus reconstruct a unique hidden trajectory $\tau_Y$. For $\gamma_Y>0$ (bottom) we still learn the states of the demon at the time of the visible transitions, but we can only probabilistically determine which hidden transitions occurred.\label{fig:trajectories}}
\end{figure}
The simplest scenario is the regime of perfect feedback, when $\gamma_X=0=\gamma_Y$. In this regime each system is only coupled to a single bath at a time. Whether it is the hot or the cold bath depends on the state of the other system. Thus changing the state of one system acts as a perfect switch for the second system's bath coupling.

In this case knowing the visible trajectory $\tau_X$ actually allows us to fully reconstruct the hidden trajectory as well (except for irrelevant transition times and transitions that cancel each other). In this case each of the possible transitions (both visible and hidden) connects a unique initial state with a unique end state, the jump operators all being of the form $L_k \propto \ket{\chi_k^f}\bra{\chi_k^i}$ (with $\ket{\chi_k^{i/f}}\in\{\ket{g,0},\ket{g,1},\ket{e,0},\ket{e,1}\}$), e.g. $L_1 \propto \ket{g,1}\bra{e,1}$. Thus we know exactly in which state each qubit was right before a visible transition and in which state it is right after the transition. If the end state of one visible transition is not the initial state of the consecutive transition, we know that a hidden transition must have occurred in between. Note that in principle any odd number of hidden transitions may have occurred, but the environmental entropy production for all of these cases is the same so we do not have to distinguish between them here. Similarly the case of no hidden transition and an even number of hidden back and forth transitions are equivalent here since they all have zero environmental entropy production in the demon system.

Given all this information we can fully and unambiguously determine the hidden trajectory $\tau_Y$ and its associated hidden entropy production as the reconstructed environmental entropy production $\Delta s_{{\rm env}}(\tau_Y)$ of the demon system, giving simply
\begin{eqnarray}\label{DeltaID_simple}
{\Delta \sigma}_Y(\tau_X) =  \Delta s_{{\rm env}}(\tau_Y),
\end{eqnarray}
without any coarse-graining taking place. The actual derivation of this result from Eq. (\ref{eq:DeltaIY}) is given in Appendix \ref{app:ExampleDerivation} and also follows as a special case of Eq. (\ref{DeltaID_medium}) presented in the next subsection.

The top half of Fig. \ref{fig:trajectories} shows the reduced state of the demon during the visible trajectory $\tau_X = \{\ket{g,0}; (k_1 = 4, t_1 = 0.9); (k_2 = 1, t_2 = 1.5); (k_3 = 4, t_3 = 2.4); (\ket{e,1},T=3)\}$. The bath temperatures here and in all the following explicit calculations were chosen to be $\beta_Y^H = 0.5$, $\beta_X^H = 1$, $\beta_X^C = 2$ and $\beta_Y^C=4$ and we ignored the direct coupling, setting $\lambda = 0$. With these bath temperatures, we find using Eqs. (\ref{eq:EnvEntX}) the environmental entropy production of the visible trajectory $\Delta s_{{\rm env}} (\tau_X) = \Delta s_4 + \Delta s_1 + \Delta s_4 = -3.0$. 

The probabilities
\begin{equation} \label{eq:TransitionProb}
p_{k_y}^j(\tau_X) = \frac{\Bigl| \braket{\chi_{k_{j+1}}^i | L_{k_y} | \chi_{k_j}^f} \Bigr|^2}{\sum_{k_y'} \Bigl| \braket{\chi_{k_{j+1}}^i | L_{k_y'} | \chi_{k_j}^f} \Bigr|^2}
\end{equation} 
shown in between visible jumps in Fig. \ref{fig:trajectories} denote the probability of the hidden transition $k_y$ having occurred in the interval between $t_{j}$ and $t_{j+1}$. Here $\ket{\chi_{k_j}^f}$ denotes the final state of the visible transition $k_j$ (with  $\ket{\chi_{k_0}^f} := \ket{a}$) and $\ket{\chi_{k_j}^i}$ being the initial state of the transition (with $\ket{\chi_{k_{M+1}}^i} := \ket{b}$). We see that in this regime of optimal feedback we know that either no or one particular transition occurred and using Eqs. (\ref{eq:EnvEntY}) can simply read off the total hidden entropy production, in this particular case ${\Delta \sigma}_Y(\tau_X) = \Delta s_6 +  \Delta s_7 + \Delta s_6 = 3.0$. 

\begin{figure}
\includegraphics[width=0.95\columnwidth]{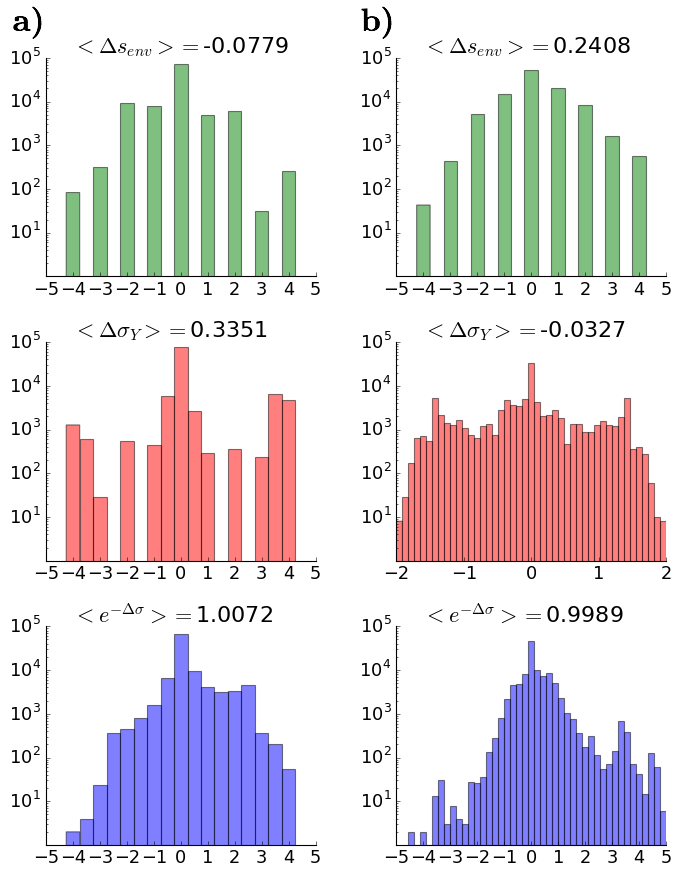}
\caption{\textbf{Histograms of entropy production.} Frequency of environmental visible entropy production  $\Delta s_{{\rm env}}$ (green, top), coarse-grained hidden entropy production ${\Delta \sigma}_Y$ (red, middle) and combined entropy production $\Delta \sigma$ (blue, bottom) for 100,000 trajectories. For $\gamma_X=0=\gamma_Y$ (\textbf{a}, left) the hidden trajectory is fully reconstructible and ${\Delta \sigma}_Y$ can only take on additive values of the demon's environmental entropy productions, whereas for $\gamma_X=0.5=\gamma_Y$ (\textbf{b}, right) the hidden entropy can have any real value. On the left $Y$ induces a negative heat flow in $X$ on average, $\braket{\Delta s_{{\rm env}}} < 0$, truly acting as a Maxwell's demon, compensating with a larger positive entropy production $\braket{{\Delta \sigma}_Y} > -\braket{\Delta s_{{\rm env}}}$. In both cases we see that the IFT is satisfied, $\braket{\exp(-\Delta \sigma)} \approx 1$.\label{fig:histograms}}
\end{figure}
The left hand side of Fig. \ref{fig:histograms} shows the distribution of visible entropy production $\Delta s_{{\rm env}} (\tau_X)$, coarse-grained hidden entropy production ${\Delta \sigma}_Y(\tau_X)$, and combined entropy production $\Delta \sigma(\tau_X)$ over 100,000 trajectories with \mbox{$T=3$} and the initial state $\ket{a}$ sampled from the steady-state distribution over the states $\{\ket{g,0},\ket{g,1},\ket{e,0},\ket{e,1}\}$. We clearly observe an average negative visible entropy production $\braket{\Delta s_{{\rm env}} (\tau_X)} < 0$, i.e. heat being extracted from the system's cold bath and transferred to its hot bath, in seeming violation of the second law. Looking at the hidden entropy production we see that this violation is more than compensated by the demon such that the modified second law Eq. (\ref{eq:SecondLaw}) which takes the feedback by the demon into account holds. We also see that the IFT Eq. (\ref{eq:IFTModified}) is satisfied as expected.

\subsection{Reconstructible Hidden States}

We next turn to the regime where $\gamma_X$ is still zero, but $\gamma_Y>0$. In this case the system jump operators (\ref{eq:LSystem}) are still of the form $L_k \propto \ket{\chi_k^f}\bra{\chi_k^i}$, connecting a unique initial state with a unique final state, but the demon transitions (\ref{eq:LDemon}) lose this definiteness, now linking multiple pairs of states. Hence, we still know the exact state of the joint system right before and after a visible transition, but if the visible trajectory reveals that hidden transitions must have taken place in between visible jumps, we can no longer be certain exactly which demon transition was responsible for the state change. If we for example consider the states $\ket{g,0}$ and $\ket{g,1}$ in Fig. \ref{fig:state}, they are now linked via hidden transitions via both of the demon's baths. Thus the feedback from the system on the demon is diminished, the system's state having a weaker influence on the demon's bath coupling. Or in other words the demon's ability to measure the state of the system diminishes, showing the complementarity of measurement by one subsystem to feedback performed by the other subsystem in autonomous systems. 

Using Eq. (\ref{eq:TransitionProb}) we can calculate the relative probability $p_{k_y}^j$ of the hidden jump $k_y$ having occurred in the $j$th interval. The bottom half of Fig. \ref{fig:trajectories} shows the reduced state of the demon and the hidden transition probabilities for the same sample trajectory $\tau_X$ as discussed in the previous section, but with $\gamma_Y = 0.5$.

In this regime the knowledge of the visible trajectory $\tau_X$ does not allow us to fully reconstruct the hidden trajectory anymore since each $\tau_X$ is compatible with multiple $\tau_Y$. We can thus no longer exactly infer the environmental entropy production in the hidden system. However, since we still precisely know the states at the times of hidden transitions and are thus able to calculate the probabilities of hidden transitions $p_{k_y}^j$ in each interval, we can use a kind of coarse-graining to make the best estimate of the hidden entropy production based on the information revealed by the visible trajectory. Specifically, we find that in this regime the general expression Eq. (\ref{eq:DeltaIY}) for ${\Delta \sigma}_Y$ reduces to
\begin{eqnarray}\label{DeltaID_medium}
{\Delta \sigma}_Y(\tau_X) =  -\sum_{j=0}^M \ln\sum_{k\in k_Y}p_{k}^j(\tau_X) e^{-\Delta s_{k}}.
\end{eqnarray}
The proof is given in Appendix \ref{app:ExampleDerivation}. We see that Eq. (\ref{DeltaID_simple}) in the previous subsection is just a special case of (\ref{DeltaID_medium}) in which all but one $p_{k}^j$ are zero during each interval $j$ so that no coarse-graining takes place. Further note that we can obtain Eq. (\ref{eq:CoarseGrained}) by converting the sum over logarithms in (\ref{DeltaID_medium}) into a single logarithm and then finding the conditional probability over full hidden trajectories $P(\tau_Y | \tau_X)$ by multiplying the relevant probabilities $p_{k}^j$ for each interval. 

Applying this result to the trajectory in Fig. \ref{fig:trajectories} we find ${\Delta \sigma}_Y(\tau_X) \approx 0.225$. Note that whereas in the case of perfect feedback ${\Delta \sigma}_Y$ could only take on additive multiples of the actual environmental entropy productions in the demon, now the incomplete information flow leads to a coarse-grained mixture of these values.

\subsection{Non-Reconstructible Hidden Trajectory}
\begin{figure}
\includegraphics[width=0.75\columnwidth]{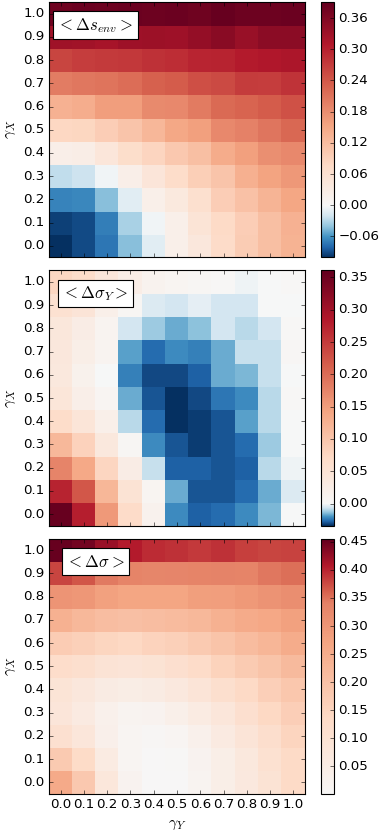}
\caption{\textbf{Average entropy production.} Average values of environmental visible entropy production $\braket{\Delta s_{{\rm env}}}$ (top), coarse-grained hidden entropy production $\braket{{\Delta \sigma}_Y}$ (middle) and combined entropy production $\braket{\Delta \sigma}$ (bottom) for different values of $\gamma_X$ and $\gamma_Y$, averaged over 50,000 trajectories each. For strong feedback, i.e. small $\gamma_X$ and $\gamma_Y$, $Y$ acts as a Maxwell's demon with $\braket{\Delta s_{{\rm env}}} < 0$, compensated by $\braket{{\Delta \sigma}_Y} > 0$. In most regions where $\braket{\Delta s_{{\rm env}}} > 0$ on the other hand, $\braket{{\Delta \sigma}_Y}$ is slightly negative which leads to Eq. (\ref{eq:SecondLaw}) giving a lower bound than the conventional second law applied to the visible system. As $\gamma_X$ and $\gamma_Y$ approach unity the feedback becomes weaker, and at unity there is no information flowing between the systems, leading to vanishing hidden entropy $\braket{{\Delta \sigma}_Y}$.\label{fig:entropy}}
\end{figure}
Let us finally consider the regime where $\gamma_X$ is also allowed to be non-zero, leading to non-optimal feedback from the demon on the system. In this case the visible transitions also connect multiple states of the joint system and the information provided by the visible trajectory only allows us to unambiguously determine the state of the system $X$, but no longer the demon $Y$. Fig. \ref{fig:trajectory} clearly demonstrates this, where we have again used the same example visible trajectory $\tau_X$ as in Fig. \ref{fig:trajectories}. Opposed to the case of $\gamma_X = 0$, the information provided by the observed jumps still leads to a back-action in the hidden system but not enough to project it into a definite pure state. Keeping the transition diagram Fig. \ref{fig:state} in mind, consider starting in the state $\ket{g,0}$. Before the first visible jump, the system may or may not undergo a hidden transition to $\ket{g,1}$. If the observed jump is now an excitation of the system with respect to its cold bath we can no longer be sure that the joint system remained in $\ket{g,0}$ before the visible jump. Thus even after the jump we find ourselves in a mixture of $\ket{e,0}$ and $\ket{e,1}$, only the visible system's state being unambiguously known. 

Further, having the intervals between visible transitions bounded by well defined states allowed us to effectively treat each interval individually in the previous cases, and the exact timing of the visible transitions was irrelevant, only their order influenced the hidden entropy production. This is no longer the case when $\gamma_X > 0$. We can no longer split the trajectory into independent intervals but have to consider the full trajectory including the precise timings to determine ${\Delta \sigma}_Y$. Considering again the scenario of starting in $\ket{g,0}$, if the visible jump happens almost immediately we can be much more certain that the hidden system did not transition to $\ket{g,1}$ in the interim than if a longer time had passed. Thus the mixture of $\ket{e,0}$ and $\ket{e,1}$ after the jump depends on the precise timing of the transition, which will further influence each future transition. 

Another example illustrates how the revealed information depends on $\gamma_X$ and the precise visible transition. Before the visible jump we know we are in a mixture of $\ket{g,0}$ and $\ket{g,1}$. Let us assume that $\gamma_X$ is small but non-zero. After an excitation of the system we will find ourselves in a mixture of $\ket{e,0}$ and $\ket{e,1}$. If the excitation was with respect to the hot (cold) bath, we know that we can give relatively more weight to $\ket{e,0}$ ($\ket{e,1}$). At the other extreme when $\gamma_X\rightarrow1$ there is no longer any feedback from the demon, and the system does not extract any information about the demon during the jump or cause any backaction on the demon state (as can also be seen by the fact that if $\gamma_X=1$, all system jump operators are of the form $L_k \propto \sigma_{\pm} \otimes \mathbb{1}$), i.e. the jump does not reveal any new information about the state of the hidden demon.

We believe that Eq. (\ref{eq:CoarseGrained}) for ${\Delta \sigma}_Y$ still holds in this regime if one can determine the correct probabilities $P(\tau_Y | \tau_X)$ but have not shown this explicitly. Trying to apply a kind of Bayesian state updating similar to \cite{Bartolotta2015} and determining the probabilities in this way did not appear to give the correct result. In all explicit calculations we have used the general (but somewhat opaque) expression Eq. (\ref{eq:DeltaIY}) which can be readily calculated and only depends on very few assumptions.

The right hand side of Fig. \ref{fig:histograms} shows the distribution of the visible, hidden and combined entropy productions for $100,000$ trajectories for $\gamma_X = 0.5 = \gamma_Y$. We note that in this regime of weaker feedback and restricted information flow, the visible entropy production is no longer negative, $\braket{\Delta s_{evn}(\tau_X)} > 0$. In fact the roles are somewhat reversed for this set of parameters with a slightly negative hidden entropy production on average. We note also that depending on the precise details of $\tau_X$, the hidden entropy can now assume any real value. Nevertheless, as expected the IFT Eq. (\ref{eq:IFTModified}) again holds with $\braket{\exp(-\Delta s_{{\rm env}}-{\Delta \sigma}_Y)} \approx 1$.

In Fig. \ref{fig:entropy} we show the average values of the three entropies for the full range of feedback parameters $\gamma_X$ and $\gamma_Y$. We see that for the given model and parameters, only for small $\gamma_X$ and $\gamma_Y$ i.e. strong measurement and feedback in both directions, the hidden system acts as a Maxwell's demon, inducing a negative entropy production in the system. In this regime the large positive hidden entropy ${\Delta \sigma}_Y$ more than compensates the negative visible entropy production. For certain intermediate values the role of demon is actually reversed with slightly negative hidden entropy production. This leads to Eq. (\ref{eq:SecondLaw}) providing a tighter bound on the entropy production than we would get from simply applying the conventional second law to the visible system. Finally as $\gamma_X,\gamma_Y\rightarrow 1$ and the systems become independent, we see that $\braket{{\Delta \sigma}_Y}$ vanishes since the visible trajectory gives us absolutely no information about the hidden system and we hence cannot make any inference regarding the hidden entropy. The bottom panel shows that the second law taking the hidden entropy into account is satisfied everywhere, $\braket{\Delta s_{{\rm env}}+{\Delta \sigma}_Y} \ge 0$. 

\begin{figure}
\includegraphics[width=\columnwidth]{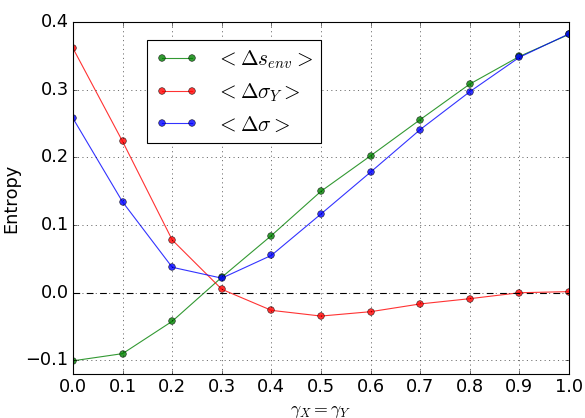}
\caption{\textbf{Two regimes of feedback.} Average values of environmental visible entropy production $\braket{\Delta s_{\rm env}}$ (green), coarse-grained hidden entropy production $\braket{{\Delta \sigma}_Y}$ (red) and combined entropy production $\braket{\Delta \sigma}$ (blue) for different values of $\gamma_X=\gamma_Y$, averaged over 50,000 trajectories each. These curves correspond to the bottom-left to top-right diagonals of the plots in Fig. \ref{fig:entropy}. We can clearly observe a split into two regions. For strong feedback $\gamma_X,\gamma_Y\lesssim0.3$ the hidden system acts as a Maxwell's demon which would lead to a violation of the second law naively applied to the visible entropy production only, $\braket{\Delta s_{\rm env}} > 0$. On the other hand for $\gamma_X,\gamma_Y\gtrsim0.3$ the hidden entropy production is negative, so that $\braket{\Delta \sigma}>0$ provides a tighter bound than the conventional second law. In the limit of no feedback, $\gamma_X,\gamma_Y\rightarrow1$, the hidden entropy vanishes and $\braket{\Delta \sigma}=\braket{\Delta s_{\rm env}}$ so that the modified and standard second laws are equivalent.\label{fig:SecondLaw}}
\end{figure}

Fig. \ref{fig:SecondLaw} shows the same results as Fig. \ref{fig:entropy}, but only for equal feedback parameters $\gamma_X = \gamma_Y$ which allows for a better comparison of the three quantities, showing even more clearly how the modified second law $\braket{\Delta \sigma} > 0$ can either explain a seeming violation of the conventional second law or provide a tighter bound on the entropy production. We have not shown this in the figures, but we also confirmed that the IFT is indeed satisfied for the full range of feedback parameters.

\subsection{Coupling of the Degenerate States}
We have so far mostly ignored the direct interaction term in the Hamiltonian proportional to $\lambda$ and have set $\lambda = 0$ for all explicit numerical results presented. Let us here briefly remark on the case of $\lambda > 0$. In the case of $\gamma_X = 0$ the simple results Eqs. (\ref{DeltaID_simple}) and (\ref{DeltaID_medium}) still hold as has been numerically verifed. Due to the diagonal split in the state diagram Fig. \ref{fig:trajectories}, the transition via $\lambda$ is equivalent to one hidden plus one visible transition, and cannot fully be achieved via hidden transitions only. By looking at the state diagram and considering the possible trajectories, one can quickly verify that even with $\lambda > 0$ we can still fully reconstruct the environmental entropy production of the demon if $\gamma_Y=0$, or use the expression (\ref{eq:TransitionProb}) to calculate the probabilities of hidden transition. Note however that while the entropy production takes the same form, $\lambda > 0$ changes the probabilities of certain trajectories and enables some previously forbidden trajectories such as $\tau_X = \{\ket{e,0}, (\ket{g,1}\rightarrow\ket{e,1} , t_1), (\ket{e,0},T)\}$. In this case the first transition would have been impossible for $\lambda = 0$ since the state $\ket{g,1}$ is not reachable from $\ket{e,0}$ without a visible transition. For $\gamma_X>0$, numerics show that the hidden entropy production associated with a given visible trajectory $\tau_X$ does depend on the value of $\lambda$, but that the dependence on $\lambda$ is very small. We have thus mostly ignored $\lambda$, also because it does not introduce any interesting new effects. Despite generating correlations, the correlations only occur within an energy subspace and are thus not particularly interesting from a thermodynamic point of view.

\section{Discussion\label{sec:Outlook}}
\begin{figure}[t!]
\includegraphics[width=0.8\columnwidth]{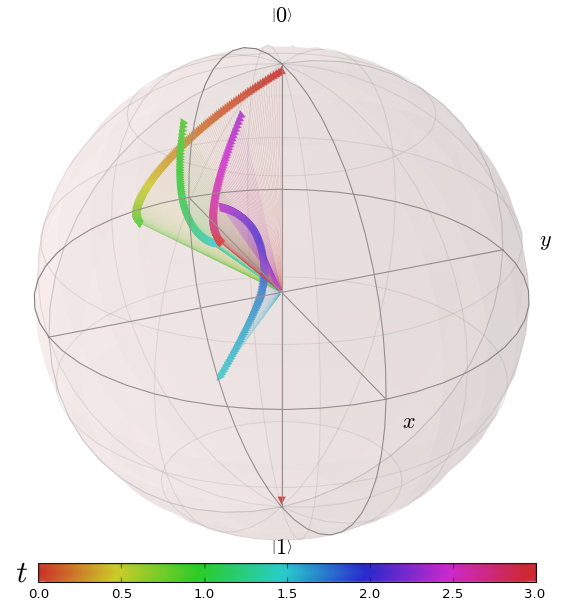}
\caption{\textbf{Fully quantum hidden evolution.} Hidden evolution of the demon $Y$ for the same visible trajectory $\tau_X$ considered in Fig. \ref{fig:trajectory}, but with an additional driving term in the demon Hamiltonian inducing coherences. Visible jumps lead to backactions in the hidden system that can change the coherence. This allows the investigation of truly quantum effects, which are captured in the hidden entropy production ${\Delta \sigma}_Y$.\label{fig:bloch}}
\end{figure}
In this article we have address the question of entropy production in autonomous quantum systems where a subset of the possible transitions is hidden and not experimentally accessible. We have shown that there is a continuous mutual measurement and feedback process taking place between the visible and hidden parts of the system that allows the hidden system to act as an autonomous Maxwell's demon. This process, which can be seen as a continuous exchange of information between the subsystems, also allows us to make inferences about the hidden system based purely on observations of the visible transitions.

In the classical setting, many articles have addressed the issue of coarse-graining \cite{Kawai2007,Esposito2012,Mehl2012,Kawaguchi2013,Bartolotta2015,Kawai2007,Sagawa2012a}. Here we have introduced a quantum \textit{coarse-grained hidden entropy production} ${\Delta \sigma}_Y$ and showed that the sum of this inferred hidden entropy and the actual observed entropy production obeys an integral fluctuation theorem. This also directly leads to a second law like inequality which can explain a negative entropy production in the visible system.

In the example presented here we still did not touch on some of the most interesting signatures of quantum thermodynamic processes such as coherences and entanglement. All the systems stayed diagonal in their energy eigenbasis during all the processes considered. However, our general results apply also to settings where this is not the case. As an example, Fig. \ref{fig:bloch} shows the evolution of the demon for the same visible trajectory $\tau_X$ and with the same parameters as in the bottom half of Fig. \ref{fig:trajectories}, but with an additional driving term $H_{d} = \mathbb{1}\otimes(\sigma_+ + \sigma_-)$ added to the Hamiltonian which induces coherences in the demon's energy eigenbasis. We see that now the visible jumps actually lead to a discontinuous back-action in the demon that can also change coherences. Using Eq. (\ref{eq:DeltaIY}) we can calculate the coarse-grained hidden entropy for this and other fully quantum cases, but a true operational understanding is still missing. This, and how the results we have presented here are related to other truly quantum mechanical thermodynamic quantities such as the quantum heat introduced in \cite{Elouard2016a}, are other promising areas for future investigations.

\section{Acknowledgments}
We acknowledge useful discussions with Terry Rudolph and David Jennings. This work was partially supported by the COST Action MP1209. MF is supported by the EPSRC and as an International Research Fellow of the Japan Society for the Promotion of Science. TS is supported by JSPS KAKENHI Grant Numbers JP16H02211 and JP25103003.

\bibliography{HiddenEntropy.bib}

\begin{thebibliography}{69}%
\makeatletter
\providecommand \@ifxundefined [1]{%
 \@ifx{#1\undefined}
}%
\providecommand \@ifnum [1]{%
 \ifnum #1\expandafter \@firstoftwo
 \else \expandafter \@secondoftwo
 \fi
}%
\providecommand \@ifx [1]{%
 \ifx #1\expandafter \@firstoftwo
 \else \expandafter \@secondoftwo
 \fi
}%
\providecommand \natexlab [1]{#1}%
\providecommand \enquote  [1]{``#1''}%
\providecommand \bibnamefont  [1]{#1}%
\providecommand \bibfnamefont [1]{#1}%
\providecommand \citenamefont [1]{#1}%
\providecommand \href@noop [0]{\@secondoftwo}%
\providecommand \href [0]{\begingroup \@sanitize@url \@href}%
\providecommand \@href[1]{\@@startlink{#1}\@@href}%
\providecommand \@@href[1]{\endgroup#1\@@endlink}%
\providecommand \@sanitize@url [0]{\catcode `\\12\catcode `\$12\catcode
  `\&12\catcode `\#12\catcode `\^12\catcode `\_12\catcode `\%12\relax}%
\providecommand \@@startlink[1]{}%
\providecommand \@@endlink[0]{}%
\providecommand \url  [0]{\begingroup\@sanitize@url \@url }%
\providecommand \@url [1]{\endgroup\@href {#1}{\urlprefix }}%
\providecommand \urlprefix  [0]{URL }%
\providecommand \Eprint [0]{\href }%
\providecommand \doibase [0]{http://dx.doi.org/}%
\providecommand \selectlanguage [0]{\@gobble}%
\providecommand \bibinfo  [0]{\@secondoftwo}%
\providecommand \bibfield  [0]{\@secondoftwo}%
\providecommand \translation [1]{[#1]}%
\providecommand \BibitemOpen [0]{}%
\providecommand \bibitemStop [0]{}%
\providecommand \bibitemNoStop [0]{.\EOS\space}%
\providecommand \EOS [0]{\spacefactor3000\relax}%
\providecommand \BibitemShut  [1]{\csname bibitem#1\endcsname}%
\let\auto@bib@innerbib\@empty
\bibitem [{\citenamefont {Maxwell}(1872)}]{maxwell1872theory}%
  \BibitemOpen
  \bibfield  {author} {\bibinfo {author} {\bibfnamefont {J.~C.}\ \bibnamefont
  {Maxwell}},\ }\href@noop {} {\emph {\bibinfo {title} {{Theory of Heat}}}},\
  Text-books of science\ (\bibinfo  {publisher} {Longmans, Green, and
  Company},\ \bibinfo {year} {1872})\BibitemShut {NoStop}%
\bibitem [{\citenamefont {Fermi}(1956)}]{Fermi1956}%
  \BibitemOpen
  \bibfield  {author} {\bibinfo {author} {\bibfnamefont {E.}~\bibnamefont
  {Fermi}},\ }\href@noop {} {\emph {\bibinfo {title} {{Thermodynamics}}}}\
  (\bibinfo  {publisher} {Dover Publications},\ \bibinfo {address} {New York},\
  \bibinfo {year} {1956})\BibitemShut {NoStop}%
\bibitem [{\citenamefont {Jarzynski}(1997)}]{Jarzynski1997}%
  \BibitemOpen
  \bibfield  {author} {\bibinfo {author} {\bibfnamefont {C.}~\bibnamefont
  {Jarzynski}},\ }\href {\doibase 10.1103/PhysRevLett.78.2690} {\bibfield
  {journal} {\bibinfo  {journal} {Physical Review Letters}\ }\textbf {\bibinfo
  {volume} {78}},\ \bibinfo {pages} {2690} (\bibinfo {year}
  {1997})}\BibitemShut {NoStop}%
\bibitem [{\citenamefont {Crooks}(1999)}]{Crooks1999_fluct}%
  \BibitemOpen
  \bibfield  {author} {\bibinfo {author} {\bibfnamefont {G.}~\bibnamefont
  {Crooks}},\ }\href {\doibase 10.1103/PhysRevE.60.2721} {\bibfield  {journal}
  {\bibinfo  {journal} {Physical Review E}\ }\textbf {\bibinfo {volume} {60}},\
  \bibinfo {pages} {2721} (\bibinfo {year} {1999})}\BibitemShut {NoStop}%
\bibitem [{\citenamefont {Lieb}\ and\ \citenamefont
  {Yngvason}(1999)}]{Lieb1999}%
  \BibitemOpen
  \bibfield  {author} {\bibinfo {author} {\bibfnamefont {E.~H.}\ \bibnamefont
  {Lieb}}\ and\ \bibinfo {author} {\bibfnamefont {J.}~\bibnamefont
  {Yngvason}},\ }\href {\doibase 10.1016/S0370-1573(98)00082-9} {\bibfield
  {journal} {\bibinfo  {journal} {Physics Reports}\ }\textbf {\bibinfo {volume}
  {310}},\ \bibinfo {pages} {1} (\bibinfo {year} {1999})}\BibitemShut {NoStop}%
\bibitem [{\citenamefont {Seifert}(2012)}]{Seifert2012}%
  \BibitemOpen
  \bibfield  {author} {\bibinfo {author} {\bibfnamefont {U.}~\bibnamefont
  {Seifert}},\ }\href {\doibase 10.1088/0034-4885/75/12/126001} {\bibfield
  {journal} {\bibinfo  {journal} {Reports on progress in physics. Physical
  Society (Great Britain)}\ }\textbf {\bibinfo {volume} {75}},\ \bibinfo
  {pages} {126001} (\bibinfo {year} {2012})}\BibitemShut {NoStop}%
\bibitem [{\citenamefont {Horowitz}\ and\ \citenamefont
  {Esposito}(2014)}]{Horowitz2014a}%
  \BibitemOpen
  \bibfield  {author} {\bibinfo {author} {\bibfnamefont {J.~M.}\ \bibnamefont
  {Horowitz}}\ and\ \bibinfo {author} {\bibfnamefont {M.}~\bibnamefont
  {Esposito}},\ }\href {\doibase 10.1103/PhysRevX.4.031015} {\bibfield
  {journal} {\bibinfo  {journal} {Physical Review X}\ }\textbf {\bibinfo
  {volume} {4}},\ \bibinfo {pages} {1} (\bibinfo {year} {2014})},\ \Eprint
  {http://arxiv.org/abs/1402.3276} {arXiv:1402.3276} \BibitemShut {NoStop}%
\bibitem [{\citenamefont {Hartich}\ \emph {et~al.}(2014)\citenamefont
  {Hartich}, \citenamefont {Barato},\ and\ \citenamefont
  {Seifert}}]{Hartich2014}%
  \BibitemOpen
  \bibfield  {author} {\bibinfo {author} {\bibfnamefont {D.}~\bibnamefont
  {Hartich}}, \bibinfo {author} {\bibfnamefont {A.~C.}\ \bibnamefont {Barato}},
  \ and\ \bibinfo {author} {\bibfnamefont {U.}~\bibnamefont {Seifert}},\ }\href
  {\doibase 10.1088/1742-5468/2014/02/P02016} {\bibfield  {journal} {\bibinfo
  {journal} {Journal of Statistical Mechanics: Theory and Experiment}\ }\textbf
  {\bibinfo {volume} {2014}},\ \bibinfo {pages} {P02016} (\bibinfo {year}
  {2014})},\ \Eprint {http://arxiv.org/abs/1402.0419} {arXiv:1402.0419}
  \BibitemShut {NoStop}%
\bibitem [{\citenamefont {Wiseman}(1996)}]{Wiseman1996}%
  \BibitemOpen
  \bibfield  {author} {\bibinfo {author} {\bibfnamefont {H.~M.}\ \bibnamefont
  {Wiseman}},\ }\href {\doibase 10.1088/1355-5111/8/1/015} {\bibfield
  {journal} {\bibinfo  {journal} {Quantum and Semiclassical Optics: Journal of
  the European Optical Society Part B}\ }\textbf {\bibinfo {volume} {8}},\
  \bibinfo {pages} {205} (\bibinfo {year} {1996})},\ \Eprint
  {http://arxiv.org/abs/0302080} {arXiv:0302080 [quant-ph]} \BibitemShut
  {NoStop}%
\bibitem [{\citenamefont {Plenio}\ and\ \citenamefont
  {Knight}(1998)}]{Plenio1998a}%
  \BibitemOpen
  \bibfield  {author} {\bibinfo {author} {\bibfnamefont {M.~B.}\ \bibnamefont
  {Plenio}}\ and\ \bibinfo {author} {\bibfnamefont {P.~L.}\ \bibnamefont
  {Knight}},\ }\href {\doibase 10.1103/RevModPhys.70.101} {\bibfield  {journal}
  {\bibinfo  {journal} {Reviews of Modern Physics}\ }\textbf {\bibinfo {volume}
  {70}},\ \bibinfo {pages} {101} (\bibinfo {year} {1998})},\ \Eprint
  {http://arxiv.org/abs/9702007} {arXiv:9702007 [quant-ph]} \BibitemShut
  {NoStop}%
\bibitem [{\citenamefont {Horowitz}(2012)}]{Horowitz2012}%
  \BibitemOpen
  \bibfield  {author} {\bibinfo {author} {\bibfnamefont {J.~M.}\ \bibnamefont
  {Horowitz}},\ }\href {\doibase 10.1103/PhysRevE.85.031110} {\bibfield
  {journal} {\bibinfo  {journal} {Physical Review E}\ }\textbf {\bibinfo
  {volume} {85}},\ \bibinfo {pages} {031110} (\bibinfo {year} {2012})},\
  \Eprint {http://arxiv.org/abs/1111.7199} {arXiv:1111.7199} \BibitemShut
  {NoStop}%
\bibitem [{\citenamefont {Hekking}\ and\ \citenamefont
  {Pekola}(2013)}]{Hekking2013}%
  \BibitemOpen
  \bibfield  {author} {\bibinfo {author} {\bibfnamefont {F.~W.~J.}\
  \bibnamefont {Hekking}}\ and\ \bibinfo {author} {\bibfnamefont {J.~P.}\
  \bibnamefont {Pekola}},\ }\href {\doibase 10.1103/PhysRevLett.111.093602}
  {\bibfield  {journal} {\bibinfo  {journal} {Physical Review Letters}\
  }\textbf {\bibinfo {volume} {111}},\ \bibinfo {pages} {093602} (\bibinfo
  {year} {2013})},\ \Eprint {http://arxiv.org/abs/1305.5207} {arXiv:1305.5207}
  \BibitemShut {NoStop}%
\bibitem [{\citenamefont {Strasberg}\ \emph {et~al.}(2013)\citenamefont
  {Strasberg}, \citenamefont {Schaller}, \citenamefont {Brandes},\ and\
  \citenamefont {Esposito}}]{Strasberg2013}%
  \BibitemOpen
  \bibfield  {author} {\bibinfo {author} {\bibfnamefont {P.}~\bibnamefont
  {Strasberg}}, \bibinfo {author} {\bibfnamefont {G.}~\bibnamefont {Schaller}},
  \bibinfo {author} {\bibfnamefont {T.}~\bibnamefont {Brandes}}, \ and\
  \bibinfo {author} {\bibfnamefont {M.}~\bibnamefont {Esposito}},\ }\href
  {http://link.aps.org/doi/10.1103/PhysRevLett.110.040601} {\bibfield
  {journal} {\bibinfo  {journal} {Physical Review Letters}\ }\textbf {\bibinfo
  {volume} {110}},\ \bibinfo {pages} {040601} (\bibinfo {year}
  {2013})}\BibitemShut {NoStop}%
\bibitem [{\citenamefont {Horowitz}\ and\ \citenamefont
  {Parrondo}(2013)}]{Horowitz2013}%
  \BibitemOpen
  \bibfield  {author} {\bibinfo {author} {\bibfnamefont {J.~M.}\ \bibnamefont
  {Horowitz}}\ and\ \bibinfo {author} {\bibfnamefont {J.~M.~R.}\ \bibnamefont
  {Parrondo}},\ }\href {\doibase 10.1088/1367-2630/15/8/085028} {\bibfield
  {journal} {\bibinfo  {journal} {New Journal of Physics}\ }\textbf {\bibinfo
  {volume} {15}},\ \bibinfo {pages} {085028} (\bibinfo {year} {2013})},\
  \Eprint {http://arxiv.org/abs/1305.6793} {arXiv:1305.6793} \BibitemShut
  {NoStop}%
\bibitem [{\citenamefont {Elouard}\ \emph {et~al.}(2015)\citenamefont
  {Elouard}, \citenamefont {Auff{\`{e}}ves},\ and\ \citenamefont
  {Clusel}}]{Elouard2015}%
  \BibitemOpen
  \bibfield  {author} {\bibinfo {author} {\bibfnamefont {C.}~\bibnamefont
  {Elouard}}, \bibinfo {author} {\bibfnamefont {A.}~\bibnamefont
  {Auff{\`{e}}ves}}, \ and\ \bibinfo {author} {\bibfnamefont {M.}~\bibnamefont
  {Clusel}},\ }\href {http://arxiv.org/abs/1507.00312} {\bibfield  {journal}
  {\bibinfo  {journal} {arXiv}\ ,\ \bibinfo {pages} {1}} (\bibinfo {year}
  {2015})},\ \Eprint {http://arxiv.org/abs/1507.00312} {arXiv:1507.00312}
  \BibitemShut {NoStop}%
\bibitem [{\citenamefont {Pigeon}\ and\ \citenamefont
  {Xuereb}(2016)}]{Pigeon2016}%
  \BibitemOpen
  \bibfield  {author} {\bibinfo {author} {\bibfnamefont {S.}~\bibnamefont
  {Pigeon}}\ and\ \bibinfo {author} {\bibfnamefont {A.}~\bibnamefont
  {Xuereb}},\ }\href {\doibase 10.1088/1742-5468/2016/06/063203} {\bibfield
  {journal} {\bibinfo  {journal} {Journal of Statistical Mechanics: Theory and
  Experiment}\ }\textbf {\bibinfo {volume} {2016}},\ \bibinfo {pages} {063203}
  (\bibinfo {year} {2016})},\ \Eprint {http://arxiv.org/abs/1602.07136}
  {arXiv:1602.07136} \BibitemShut {NoStop}%
\bibitem [{\citenamefont {del Rio}\ \emph {et~al.}(2011)\citenamefont {del
  Rio}, \citenamefont {Aberg}, \citenamefont {Renner}, \citenamefont
  {Dahlsten},\ and\ \citenamefont {Vedral}}]{Renner2011}%
  \BibitemOpen
  \bibfield  {author} {\bibinfo {author} {\bibfnamefont {L.}~\bibnamefont {del
  Rio}}, \bibinfo {author} {\bibfnamefont {J.}~\bibnamefont {Aberg}}, \bibinfo
  {author} {\bibfnamefont {R.}~\bibnamefont {Renner}}, \bibinfo {author}
  {\bibfnamefont {O.}~\bibnamefont {Dahlsten}}, \ and\ \bibinfo {author}
  {\bibfnamefont {V.}~\bibnamefont {Vedral}},\ }\href {\doibase
  10.1038/nature10123} {\bibfield  {journal} {\bibinfo  {journal} {Nature}\
  }\textbf {\bibinfo {volume} {474}},\ \bibinfo {pages} {61} (\bibinfo {year}
  {2011})},\ \Eprint {http://arxiv.org/abs/1009.1630} {arXiv:1009.1630}
  \BibitemShut {NoStop}%
\bibitem [{\citenamefont {Aberg}(2013)}]{Aberg2011}%
  \BibitemOpen
  \bibfield  {author} {\bibinfo {author} {\bibfnamefont {J.}~\bibnamefont
  {Aberg}},\ }\href {\doibase 10.1038/ncomms2712} {\bibfield  {journal}
  {\bibinfo  {journal} {Nature communications}\ }\textbf {\bibinfo {volume}
  {4}},\ \bibinfo {pages} {1925} (\bibinfo {year} {2013})},\ \Eprint
  {http://arxiv.org/abs/1110.6121} {arXiv:1110.6121} \BibitemShut {NoStop}%
\bibitem [{\citenamefont {Horodecki}\ and\ \citenamefont
  {Oppenheim}(2013)}]{Horodecki2011}%
  \BibitemOpen
  \bibfield  {author} {\bibinfo {author} {\bibfnamefont {M.}~\bibnamefont
  {Horodecki}}\ and\ \bibinfo {author} {\bibfnamefont {J.}~\bibnamefont
  {Oppenheim}},\ }\href {\doibase 10.1038/ncomms3059} {\bibfield  {journal}
  {\bibinfo  {journal} {Nature communications}\ }\textbf {\bibinfo {volume}
  {4}},\ \bibinfo {pages} {2059} (\bibinfo {year} {2013})},\ \Eprint
  {http://arxiv.org/abs/1111.3834} {arXiv:1111.3834} \BibitemShut {NoStop}%
\bibitem [{\citenamefont {Leggio}\ \emph {et~al.}(2013)\citenamefont {Leggio},
  \citenamefont {Napoli}, \citenamefont {Messina},\ and\ \citenamefont
  {Breuer}}]{Leggio2013}%
  \BibitemOpen
  \bibfield  {author} {\bibinfo {author} {\bibfnamefont {B.}~\bibnamefont
  {Leggio}}, \bibinfo {author} {\bibfnamefont {A.}~\bibnamefont {Napoli}},
  \bibinfo {author} {\bibfnamefont {A.}~\bibnamefont {Messina}}, \ and\
  \bibinfo {author} {\bibfnamefont {H.-P.}\ \bibnamefont {Breuer}},\ }\href
  {\doibase 10.1103/PhysRevA.88.042111} {\bibfield  {journal} {\bibinfo
  {journal} {Physical Review A}\ }\textbf {\bibinfo {volume} {88}},\ \bibinfo
  {pages} {042111} (\bibinfo {year} {2013})},\ \Eprint
  {http://arxiv.org/abs/1305.6733} {arXiv:1305.6733} \BibitemShut {NoStop}%
\bibitem [{\citenamefont {Brand{\~{a}}o}\ \emph {et~al.}(2015)\citenamefont
  {Brand{\~{a}}o}, \citenamefont {Horodecki}, \citenamefont {Ng}, \citenamefont
  {Oppenheim},\ and\ \citenamefont {Wehner}}]{Brandao2013}%
  \BibitemOpen
  \bibfield  {author} {\bibinfo {author} {\bibfnamefont {F.}~\bibnamefont
  {Brand{\~{a}}o}}, \bibinfo {author} {\bibfnamefont {M.}~\bibnamefont
  {Horodecki}}, \bibinfo {author} {\bibfnamefont {N.}~\bibnamefont {Ng}},
  \bibinfo {author} {\bibfnamefont {J.}~\bibnamefont {Oppenheim}}, \ and\
  \bibinfo {author} {\bibfnamefont {S.}~\bibnamefont {Wehner}},\ }\href
  {\doibase 10.1073/pnas.1411728112} {\bibfield  {journal} {\bibinfo  {journal}
  {Proceedings of the National Academy of Sciences}\ }\textbf {\bibinfo
  {volume} {112}},\ \bibinfo {pages} {3275} (\bibinfo {year} {2015})},\ \Eprint
  {http://arxiv.org/abs/1305.5278} {arXiv:1305.5278} \BibitemShut {NoStop}%
\bibitem [{\citenamefont {Korzekwa}\ \emph {et~al.}(2016)\citenamefont
  {Korzekwa}, \citenamefont {Lostaglio}, \citenamefont {Oppenheim},\ and\
  \citenamefont {Jennings}}]{Korzekwa}%
  \BibitemOpen
  \bibfield  {author} {\bibinfo {author} {\bibfnamefont {K.}~\bibnamefont
  {Korzekwa}}, \bibinfo {author} {\bibfnamefont {M.}~\bibnamefont {Lostaglio}},
  \bibinfo {author} {\bibfnamefont {J.}~\bibnamefont {Oppenheim}}, \ and\
  \bibinfo {author} {\bibfnamefont {D.}~\bibnamefont {Jennings}},\ }\href
  {\doibase 10.1088/1367-2630/18/2/023045} {\bibfield  {journal} {\bibinfo
  {journal} {New Journal of Physics}\ }\textbf {\bibinfo {volume} {18}},\
  \bibinfo {pages} {023045} (\bibinfo {year} {2016})},\ \Eprint
  {http://arxiv.org/abs/1506.07875} {arXiv:1506.07875} \BibitemShut {NoStop}%
\bibitem [{\citenamefont {Lostaglio}\ \emph
  {et~al.}(2015{\natexlab{a}})\citenamefont {Lostaglio}, \citenamefont
  {Jennings},\ and\ \citenamefont {Rudolph}}]{Lostaglio2014}%
  \BibitemOpen
  \bibfield  {author} {\bibinfo {author} {\bibfnamefont {M.}~\bibnamefont
  {Lostaglio}}, \bibinfo {author} {\bibfnamefont {D.}~\bibnamefont {Jennings}},
  \ and\ \bibinfo {author} {\bibfnamefont {T.}~\bibnamefont {Rudolph}},\ }\href
  {\doibase 10.1038/ncomms7383} {\bibfield  {journal} {\bibinfo  {journal}
  {Nature Communications}\ }\textbf {\bibinfo {volume} {6}},\ \bibinfo {pages}
  {6383} (\bibinfo {year} {2015}{\natexlab{a}})},\ \Eprint
  {http://arxiv.org/abs/1405.2188} {arXiv:1405.2188} \BibitemShut {NoStop}%
\bibitem [{\citenamefont {Lostaglio}\ \emph
  {et~al.}(2015{\natexlab{b}})\citenamefont {Lostaglio}, \citenamefont
  {Korzekwa}, \citenamefont {Jennings},\ and\ \citenamefont
  {Rudolph}}]{Lostaglio2015}%
  \BibitemOpen
  \bibfield  {author} {\bibinfo {author} {\bibfnamefont {M.}~\bibnamefont
  {Lostaglio}}, \bibinfo {author} {\bibfnamefont {K.}~\bibnamefont {Korzekwa}},
  \bibinfo {author} {\bibfnamefont {D.}~\bibnamefont {Jennings}}, \ and\
  \bibinfo {author} {\bibfnamefont {T.}~\bibnamefont {Rudolph}},\ }\href
  {\doibase 10.1103/PhysRevX.5.021001} {\bibfield  {journal} {\bibinfo
  {journal} {Physical Review X}\ }\textbf {\bibinfo {volume} {5}},\ \bibinfo
  {pages} {021001} (\bibinfo {year} {2015}{\natexlab{b}})}\BibitemShut
  {NoStop}%
\bibitem [{\citenamefont {Kammerlander}\ and\ \citenamefont
  {Anders}(2016)}]{Anders}%
  \BibitemOpen
  \bibfield  {author} {\bibinfo {author} {\bibfnamefont {P.}~\bibnamefont
  {Kammerlander}}\ and\ \bibinfo {author} {\bibfnamefont {J.}~\bibnamefont
  {Anders}},\ }\href {\doibase 10.1038/srep22174} {\bibfield  {journal}
  {\bibinfo  {journal} {Scientific Reports}\ }\textbf {\bibinfo {volume} {6}},\
  \bibinfo {pages} {22174} (\bibinfo {year} {2016})},\ \Eprint
  {http://arxiv.org/abs/1502.02673} {arXiv:1502.02673} \BibitemShut {NoStop}%
\bibitem [{\citenamefont {Elouard}\ \emph {et~al.}(2016)\citenamefont
  {Elouard}, \citenamefont {Herrera-Mart{\'{i}}}, \citenamefont {Clusel},\ and\
  \citenamefont {Auff{\`{e}}ves}}]{Elouard2016a}%
  \BibitemOpen
  \bibfield  {author} {\bibinfo {author} {\bibfnamefont {C.}~\bibnamefont
  {Elouard}}, \bibinfo {author} {\bibfnamefont {D.}~\bibnamefont
  {Herrera-Mart{\'{i}}}}, \bibinfo {author} {\bibfnamefont {M.}~\bibnamefont
  {Clusel}}, \ and\ \bibinfo {author} {\bibfnamefont {A.}~\bibnamefont
  {Auff{\`{e}}ves}},\ }\href {http://arxiv.org/abs/1607.02404} {\ ,\ \bibinfo
  {pages} {1} (\bibinfo {year} {2016})},\ \Eprint
  {http://arxiv.org/abs/1607.02404} {arXiv:1607.02404} \BibitemShut {NoStop}%
\bibitem [{\citenamefont {Jevtic}\ \emph {et~al.}(2015)\citenamefont {Jevtic},
  \citenamefont {Rudolph}, \citenamefont {Jennings}, \citenamefont {Hirono},
  \citenamefont {Nakayama},\ and\ \citenamefont {Murao}}]{Jevtic2015}%
  \BibitemOpen
  \bibfield  {author} {\bibinfo {author} {\bibfnamefont {S.}~\bibnamefont
  {Jevtic}}, \bibinfo {author} {\bibfnamefont {T.}~\bibnamefont {Rudolph}},
  \bibinfo {author} {\bibfnamefont {D.}~\bibnamefont {Jennings}}, \bibinfo
  {author} {\bibfnamefont {Y.}~\bibnamefont {Hirono}}, \bibinfo {author}
  {\bibfnamefont {S.}~\bibnamefont {Nakayama}}, \ and\ \bibinfo {author}
  {\bibfnamefont {M.}~\bibnamefont {Murao}},\ }\href {\doibase
  10.1103/PhysRevE.92.042113} {\bibfield  {journal} {\bibinfo  {journal}
  {Physical Review E}\ }\textbf {\bibinfo {volume} {92}},\ \bibinfo {pages}
  {042113} (\bibinfo {year} {2015})}\BibitemShut {NoStop}%
\bibitem [{\citenamefont {Browne}\ \emph {et~al.}(2016)\citenamefont {Browne},
  \citenamefont {Farrow}, \citenamefont {Dahlsten},\ and\ \citenamefont
  {Vedral}}]{Browne2016}%
  \BibitemOpen
  \bibfield  {author} {\bibinfo {author} {\bibfnamefont {C.}~\bibnamefont
  {Browne}}, \bibinfo {author} {\bibfnamefont {T.}~\bibnamefont {Farrow}},
  \bibinfo {author} {\bibfnamefont {O.~C.~O.}\ \bibnamefont {Dahlsten}}, \ and\
  \bibinfo {author} {\bibfnamefont {V.}~\bibnamefont {Vedral}},\ }\href
  {http://arxiv.org/abs/1606.03318} {\ ,\ \bibinfo {pages} {1} (\bibinfo {year}
  {2016})},\ \Eprint {http://arxiv.org/abs/1606.03318} {arXiv:1606.03318}
  \BibitemShut {NoStop}%
\bibitem [{\citenamefont {Horowitz}\ and\ \citenamefont
  {Vaikuntanathan}(2010)}]{Horowitz2010}%
  \BibitemOpen
  \bibfield  {author} {\bibinfo {author} {\bibfnamefont {J.~M.}\ \bibnamefont
  {Horowitz}}\ and\ \bibinfo {author} {\bibfnamefont {S.}~\bibnamefont
  {Vaikuntanathan}},\ }\href {\doibase 10.1103/PhysRevE.82.061120} {\bibfield
  {journal} {\bibinfo  {journal} {Physical Review E}\ }\textbf {\bibinfo
  {volume} {82}},\ \bibinfo {pages} {061120} (\bibinfo {year}
  {2010})}\BibitemShut {NoStop}%
\bibitem [{\citenamefont {Sagawa}(2011)}]{Sagawa2011}%
  \BibitemOpen
  \bibfield  {author} {\bibinfo {author} {\bibfnamefont {T.}~\bibnamefont
  {Sagawa}},\ }\href {\doibase 10.1088/1742-6596/297/1/012015} {\bibfield
  {journal} {\bibinfo  {journal} {Journal of Physics: Conference Series}\
  }\textbf {\bibinfo {volume} {297}},\ \bibinfo {pages} {012015} (\bibinfo
  {year} {2011})}\BibitemShut {NoStop}%
\bibitem [{\citenamefont {Ashida}\ \emph {et~al.}(2014)\citenamefont {Ashida},
  \citenamefont {Funo}, \citenamefont {Murashita},\ and\ \citenamefont
  {Ueda}}]{Ashida2014}%
  \BibitemOpen
  \bibfield  {author} {\bibinfo {author} {\bibfnamefont {Y.}~\bibnamefont
  {Ashida}}, \bibinfo {author} {\bibfnamefont {K.}~\bibnamefont {Funo}},
  \bibinfo {author} {\bibfnamefont {Y.}~\bibnamefont {Murashita}}, \ and\
  \bibinfo {author} {\bibfnamefont {M.}~\bibnamefont {Ueda}},\ }\href {\doibase
  10.1103/PhysRevE.90.052125} {\bibfield  {journal} {\bibinfo  {journal}
  {Physical Review E}\ }\textbf {\bibinfo {volume} {90}},\ \bibinfo {pages}
  {052125} (\bibinfo {year} {2014})},\ \Eprint {http://arxiv.org/abs/1404.2388}
  {arXiv:1404.2388} \BibitemShut {NoStop}%
\bibitem [{\citenamefont {Sagawa}\ and\ \citenamefont
  {Ueda}(2012)}]{Sagawa2012a}%
  \BibitemOpen
  \bibfield  {author} {\bibinfo {author} {\bibfnamefont {T.}~\bibnamefont
  {Sagawa}}\ and\ \bibinfo {author} {\bibfnamefont {M.}~\bibnamefont {Ueda}},\
  }\href {\doibase 10.1103/PhysRevE.85.021104} {\bibfield  {journal} {\bibinfo
  {journal} {Physical Review E}\ }\textbf {\bibinfo {volume} {85}},\ \bibinfo
  {pages} {021104} (\bibinfo {year} {2012})}\BibitemShut {NoStop}%
\bibitem [{\citenamefont {Horowitz}\ and\ \citenamefont
  {Parrondo}(2011)}]{Horowitz2011}%
  \BibitemOpen
  \bibfield  {author} {\bibinfo {author} {\bibfnamefont {J.~M.}\ \bibnamefont
  {Horowitz}}\ and\ \bibinfo {author} {\bibfnamefont {J.~M.~R.}\ \bibnamefont
  {Parrondo}},\ }\href {\doibase 10.1209/0295-5075/95/10005} {\bibfield
  {journal} {\bibinfo  {journal} {EPL (Europhysics Letters)}\ }\textbf
  {\bibinfo {volume} {95}},\ \bibinfo {pages} {10005} (\bibinfo {year}
  {2011})}\BibitemShut {NoStop}%
\bibitem [{\citenamefont {Ito}\ and\ \citenamefont {Sagawa}(2013)}]{Ito2013}%
  \BibitemOpen
  \bibfield  {author} {\bibinfo {author} {\bibfnamefont {S.}~\bibnamefont
  {Ito}}\ and\ \bibinfo {author} {\bibfnamefont {T.}~\bibnamefont {Sagawa}},\
  }\href {\doibase 10.1103/PhysRevLett.111.180603} {\bibfield  {journal}
  {\bibinfo  {journal} {Physical Review Letters}\ }\textbf {\bibinfo {volume}
  {111}},\ \bibinfo {pages} {180603} (\bibinfo {year} {2013})}\BibitemShut
  {NoStop}%
\bibitem [{\citenamefont {Funo}\ \emph {et~al.}(2013)\citenamefont {Funo},
  \citenamefont {Watanabe},\ and\ \citenamefont {Ueda}}]{Funo2013}%
  \BibitemOpen
  \bibfield  {author} {\bibinfo {author} {\bibfnamefont {K.}~\bibnamefont
  {Funo}}, \bibinfo {author} {\bibfnamefont {Y.}~\bibnamefont {Watanabe}}, \
  and\ \bibinfo {author} {\bibfnamefont {M.}~\bibnamefont {Ueda}},\ }\href
  {\doibase 10.1103/PhysRevE.88.052121} {\bibfield  {journal} {\bibinfo
  {journal} {Physical Review E}\ }\textbf {\bibinfo {volume} {88}},\ \bibinfo
  {pages} {052121} (\bibinfo {year} {2013})}\BibitemShut {NoStop}%
\bibitem [{\citenamefont {Horowitz}\ and\ \citenamefont
  {Jacobs}(2014)}]{Horowitz2014}%
  \BibitemOpen
  \bibfield  {author} {\bibinfo {author} {\bibfnamefont {J.~M.}\ \bibnamefont
  {Horowitz}}\ and\ \bibinfo {author} {\bibfnamefont {K.}~\bibnamefont
  {Jacobs}},\ }\href {\doibase 10.1103/PhysRevE.89.042134} {\bibfield
  {journal} {\bibinfo  {journal} {Physical Review E}\ }\textbf {\bibinfo
  {volume} {89}},\ \bibinfo {pages} {042134} (\bibinfo {year}
  {2014})}\BibitemShut {NoStop}%
\bibitem [{\citenamefont {Jacobs}(2009)}]{Jacobs2009}%
  \BibitemOpen
  \bibfield  {author} {\bibinfo {author} {\bibfnamefont {K.}~\bibnamefont
  {Jacobs}},\ }\href {\doibase 10.1103/PhysRevA.80.012322} {\bibfield
  {journal} {\bibinfo  {journal} {Physical Review A}\ }\textbf {\bibinfo
  {volume} {80}},\ \bibinfo {pages} {012322} (\bibinfo {year}
  {2009})}\BibitemShut {NoStop}%
\bibitem [{\citenamefont {Sagawa}\ and\ \citenamefont
  {Ueda}(2008)}]{Sagawa2008}%
  \BibitemOpen
  \bibfield  {author} {\bibinfo {author} {\bibfnamefont {T.}~\bibnamefont
  {Sagawa}}\ and\ \bibinfo {author} {\bibfnamefont {M.}~\bibnamefont {Ueda}},\
  }\href {\doibase 10.1103/PhysRevLett.100.080403} {\bibfield  {journal}
  {\bibinfo  {journal} {Physical Review Letters}\ }\textbf {\bibinfo {volume}
  {100}},\ \bibinfo {pages} {080403} (\bibinfo {year} {2008})}\BibitemShut
  {NoStop}%
\bibitem [{\citenamefont {Gong}\ \emph {et~al.}(2016)\citenamefont {Gong},
  \citenamefont {Ashida},\ and\ \citenamefont {Ueda}}]{Gong2016}%
  \BibitemOpen
  \bibfield  {author} {\bibinfo {author} {\bibfnamefont {Z.}~\bibnamefont
  {Gong}}, \bibinfo {author} {\bibfnamefont {Y.}~\bibnamefont {Ashida}}, \ and\
  \bibinfo {author} {\bibfnamefont {M.}~\bibnamefont {Ueda}},\ }\href {\doibase
  10.1103/PhysRevA.94.012107} {\bibfield  {journal} {\bibinfo  {journal}
  {Physical Review A}\ }\textbf {\bibinfo {volume} {94}},\ \bibinfo {pages}
  {012107} (\bibinfo {year} {2016})},\ \Eprint
  {http://arxiv.org/abs/1602.04616} {arXiv:1602.04616} \BibitemShut {NoStop}%
\bibitem [{\citenamefont {Frenzel}\ \emph {et~al.}(2014)\citenamefont
  {Frenzel}, \citenamefont {Jennings},\ and\ \citenamefont
  {Rudolph}}]{Frenzel2014d}%
  \BibitemOpen
  \bibfield  {author} {\bibinfo {author} {\bibfnamefont {M.~F.}\ \bibnamefont
  {Frenzel}}, \bibinfo {author} {\bibfnamefont {D.}~\bibnamefont {Jennings}}, \
  and\ \bibinfo {author} {\bibfnamefont {T.}~\bibnamefont {Rudolph}},\ }\href
  {\doibase 10.1103/PhysRevE.90.052136} {\bibfield  {journal} {\bibinfo
  {journal} {Physical Review E}\ }\textbf {\bibinfo {volume} {90}},\ \bibinfo
  {pages} {052136} (\bibinfo {year} {2014})}\BibitemShut {NoStop}%
\bibitem [{\citenamefont {Frenzel}\ \emph {et~al.}(2016)\citenamefont
  {Frenzel}, \citenamefont {Jennings},\ and\ \citenamefont
  {Rudolph}}]{Frenzel2015}%
  \BibitemOpen
  \bibfield  {author} {\bibinfo {author} {\bibfnamefont {M.~F.}\ \bibnamefont
  {Frenzel}}, \bibinfo {author} {\bibfnamefont {D.}~\bibnamefont {Jennings}}, \
  and\ \bibinfo {author} {\bibfnamefont {T.}~\bibnamefont {Rudolph}},\ }\href
  {\doibase 10.1088/1367-2630/18/2/023037} {\bibfield  {journal} {\bibinfo
  {journal} {New Journal of Physics}\ }\textbf {\bibinfo {volume} {18}},\
  \bibinfo {pages} {023037} (\bibinfo {year} {2016})},\ \Eprint
  {http://arxiv.org/abs/1508.02720} {arXiv:1508.02720} \BibitemShut {NoStop}%
\bibitem [{\citenamefont {Malabarba}\ \emph {et~al.}(2015)\citenamefont
  {Malabarba}, \citenamefont {Short},\ and\ \citenamefont
  {Kammerlander}}]{Malabarba2014a}%
  \BibitemOpen
  \bibfield  {author} {\bibinfo {author} {\bibfnamefont {A.~S.~L.}\
  \bibnamefont {Malabarba}}, \bibinfo {author} {\bibfnamefont {A.~J.}\
  \bibnamefont {Short}}, \ and\ \bibinfo {author} {\bibfnamefont
  {P.}~\bibnamefont {Kammerlander}},\ }\href {\doibase
  10.1088/1367-2630/17/4/045027} {\bibfield  {journal} {\bibinfo  {journal}
  {New Journal of Physics}\ }\textbf {\bibinfo {volume} {17}},\ \bibinfo
  {pages} {045027} (\bibinfo {year} {2015})},\ \Eprint
  {http://arxiv.org/abs/1412.1338} {arXiv:1412.1338} \BibitemShut {NoStop}%
\bibitem [{\citenamefont {Silva}\ \emph {et~al.}(2016)\citenamefont {Silva},
  \citenamefont {Manzano}, \citenamefont {Skrzypczyk},\ and\ \citenamefont
  {Brunner}}]{Silva2016}%
  \BibitemOpen
  \bibfield  {author} {\bibinfo {author} {\bibfnamefont {R.}~\bibnamefont
  {Silva}}, \bibinfo {author} {\bibfnamefont {G.}~\bibnamefont {Manzano}},
  \bibinfo {author} {\bibfnamefont {P.}~\bibnamefont {Skrzypczyk}}, \ and\
  \bibinfo {author} {\bibfnamefont {N.}~\bibnamefont {Brunner}},\ }\href
  {http://arxiv.org/abs/1604.04098} {\ ,\ \bibinfo {pages} {1} (\bibinfo {year}
  {2016})},\ \Eprint {http://arxiv.org/abs/1604.04098} {arXiv:1604.04098}
  \BibitemShut {NoStop}%
\bibitem [{\citenamefont {Shiraishi}\ and\ \citenamefont
  {Sagawa}(2015)}]{Shiraishi2015b}%
  \BibitemOpen
  \bibfield  {author} {\bibinfo {author} {\bibfnamefont {N.}~\bibnamefont
  {Shiraishi}}\ and\ \bibinfo {author} {\bibfnamefont {T.}~\bibnamefont
  {Sagawa}},\ }\href {\doibase 10.1103/PhysRevE.91.012130} {\bibfield
  {journal} {\bibinfo  {journal} {Physical Review E}\ }\textbf {\bibinfo
  {volume} {91}},\ \bibinfo {pages} {012130} (\bibinfo {year}
  {2015})}\BibitemShut {NoStop}%
\bibitem [{\citenamefont {Shiraishi}\ \emph {et~al.}(2015)\citenamefont
  {Shiraishi}, \citenamefont {Ito}, \citenamefont {Kawaguchi},\ and\
  \citenamefont {Sagawa}}]{Shiraishi2015}%
  \BibitemOpen
  \bibfield  {author} {\bibinfo {author} {\bibfnamefont {N.}~\bibnamefont
  {Shiraishi}}, \bibinfo {author} {\bibfnamefont {S.}~\bibnamefont {Ito}},
  \bibinfo {author} {\bibfnamefont {K.}~\bibnamefont {Kawaguchi}}, \ and\
  \bibinfo {author} {\bibfnamefont {T.}~\bibnamefont {Sagawa}},\ }\href
  {\doibase 10.1088/1367-2630/17/4/045012} {\bibfield  {journal} {\bibinfo
  {journal} {New Journal of Physics}\ }\textbf {\bibinfo {volume} {17}},\
  \bibinfo {pages} {045012} (\bibinfo {year} {2015})},\ \Eprint
  {http://arxiv.org/abs/1501.06071} {arXiv:1501.06071} \BibitemShut {NoStop}%
\bibitem [{\citenamefont {Camati}\ \emph {et~al.}(2016)\citenamefont {Camati},
  \citenamefont {Peterson}, \citenamefont {Batalh{\~{a}}o}, \citenamefont
  {Micadei}, \citenamefont {Souza}, \citenamefont {Sarthour}, \citenamefont
  {Oliveira},\ and\ \citenamefont {Serra}}]{Camati2016}%
  \BibitemOpen
  \bibfield  {author} {\bibinfo {author} {\bibfnamefont {P.~A.}\ \bibnamefont
  {Camati}}, \bibinfo {author} {\bibfnamefont {J.~P.~S.}\ \bibnamefont
  {Peterson}}, \bibinfo {author} {\bibfnamefont {T.~B.}\ \bibnamefont
  {Batalh{\~{a}}o}}, \bibinfo {author} {\bibfnamefont {K.}~\bibnamefont
  {Micadei}}, \bibinfo {author} {\bibfnamefont {A.~M.}\ \bibnamefont {Souza}},
  \bibinfo {author} {\bibfnamefont {R.~S.}\ \bibnamefont {Sarthour}}, \bibinfo
  {author} {\bibfnamefont {I.~S.}\ \bibnamefont {Oliveira}}, \ and\ \bibinfo
  {author} {\bibfnamefont {R.~M.}\ \bibnamefont {Serra}},\ }\href
  {http://arxiv.org/abs/1605.08821} {\ ,\ \bibinfo {pages} {1} (\bibinfo {year}
  {2016})},\ \Eprint {http://arxiv.org/abs/1605.08821} {arXiv:1605.08821}
  \BibitemShut {NoStop}%
\bibitem [{\citenamefont {Bub}(2001)}]{Bub2001}%
  \BibitemOpen
  \bibfield  {author} {\bibinfo {author} {\bibfnamefont {J.}~\bibnamefont
  {Bub}},\ }\href {\doibase 10.1016/S1355-2198(01)00023-5} {\bibfield
  {journal} {\bibinfo  {journal} {Studies in History and Philosophy of Science
  Part B: Studies in History and Philosophy of Modern Physics}\ }\textbf
  {\bibinfo {volume} {32}},\ \bibinfo {pages} {569} (\bibinfo {year} {2001})},\
  \Eprint {http://arxiv.org/abs/0203017} {arXiv:0203017 [quant-ph]}
  \BibitemShut {NoStop}%
\bibitem [{\citenamefont {Maruyama}\ \emph {et~al.}(2009)\citenamefont
  {Maruyama}, \citenamefont {Nori},\ and\ \citenamefont
  {Vedral}}]{Maruyama2009}%
  \BibitemOpen
  \bibfield  {author} {\bibinfo {author} {\bibfnamefont {K.}~\bibnamefont
  {Maruyama}}, \bibinfo {author} {\bibfnamefont {F.}~\bibnamefont {Nori}}, \
  and\ \bibinfo {author} {\bibfnamefont {V.}~\bibnamefont {Vedral}},\ }\href
  {\doibase 10.1103/RevModPhys.81.1} {\bibfield  {journal} {\bibinfo  {journal}
  {Reviews of Modern Physics}\ }\textbf {\bibinfo {volume} {81}},\ \bibinfo
  {pages} {1} (\bibinfo {year} {2009})}\BibitemShut {NoStop}%
\bibitem [{\citenamefont {Chapman}\ and\ \citenamefont
  {Miyake}(2015)}]{Chapman2015}%
  \BibitemOpen
  \bibfield  {author} {\bibinfo {author} {\bibfnamefont {A.}~\bibnamefont
  {Chapman}}\ and\ \bibinfo {author} {\bibfnamefont {A.}~\bibnamefont
  {Miyake}},\ }\href {\doibase 10.1103/PhysRevE.92.062125} {\bibfield
  {journal} {\bibinfo  {journal} {Physical Review E}\ }\textbf {\bibinfo
  {volume} {92}},\ \bibinfo {pages} {062125} (\bibinfo {year} {2015})},\
  \Eprint {http://arxiv.org/abs/1506.09207} {arXiv:1506.09207} \BibitemShut
  {NoStop}%
\bibitem [{\citenamefont {Gleyzes}\ \emph {et~al.}(2007)\citenamefont
  {Gleyzes}, \citenamefont {Kuhr}, \citenamefont {Guerlin}, \citenamefont
  {Bernu}, \citenamefont {Del{\'{e}}glise}, \citenamefont {{Busk Hoff}},
  \citenamefont {Brune}, \citenamefont {Raimond},\ and\ \citenamefont
  {Haroche}}]{Kuhr2007}%
  \BibitemOpen
  \bibfield  {author} {\bibinfo {author} {\bibfnamefont {S.}~\bibnamefont
  {Gleyzes}}, \bibinfo {author} {\bibfnamefont {S.}~\bibnamefont {Kuhr}},
  \bibinfo {author} {\bibfnamefont {C.}~\bibnamefont {Guerlin}}, \bibinfo
  {author} {\bibfnamefont {J.}~\bibnamefont {Bernu}}, \bibinfo {author}
  {\bibfnamefont {S.}~\bibnamefont {Del{\'{e}}glise}}, \bibinfo {author}
  {\bibfnamefont {U.}~\bibnamefont {{Busk Hoff}}}, \bibinfo {author}
  {\bibfnamefont {M.}~\bibnamefont {Brune}}, \bibinfo {author} {\bibfnamefont
  {J.-M.}\ \bibnamefont {Raimond}}, \ and\ \bibinfo {author} {\bibfnamefont
  {S.}~\bibnamefont {Haroche}},\ }\href {\doibase 10.1038/nature05589}
  {\bibfield  {journal} {\bibinfo  {journal} {Nature}\ }\textbf {\bibinfo
  {volume} {446}},\ \bibinfo {pages} {297} (\bibinfo {year} {2007})},\ \Eprint
  {http://arxiv.org/abs/0612031} {arXiv:0612031 [quant-ph]} \BibitemShut
  {NoStop}%
\bibitem [{\citenamefont {Barreiro}\ \emph {et~al.}(2011)\citenamefont
  {Barreiro}, \citenamefont {M{\"{u}}ller}, \citenamefont {Schindler},
  \citenamefont {Nigg}, \citenamefont {Monz}, \citenamefont {Chwalla},
  \citenamefont {Hennrich}, \citenamefont {Roos}, \citenamefont {Zoller},\ and\
  \citenamefont {Blatt}}]{Barreiro2011}%
  \BibitemOpen
  \bibfield  {author} {\bibinfo {author} {\bibfnamefont {J.~T.}\ \bibnamefont
  {Barreiro}}, \bibinfo {author} {\bibfnamefont {M.}~\bibnamefont
  {M{\"{u}}ller}}, \bibinfo {author} {\bibfnamefont {P.}~\bibnamefont
  {Schindler}}, \bibinfo {author} {\bibfnamefont {D.}~\bibnamefont {Nigg}},
  \bibinfo {author} {\bibfnamefont {T.}~\bibnamefont {Monz}}, \bibinfo {author}
  {\bibfnamefont {M.}~\bibnamefont {Chwalla}}, \bibinfo {author} {\bibfnamefont
  {M.}~\bibnamefont {Hennrich}}, \bibinfo {author} {\bibfnamefont {C.~F.}\
  \bibnamefont {Roos}}, \bibinfo {author} {\bibfnamefont {P.}~\bibnamefont
  {Zoller}}, \ and\ \bibinfo {author} {\bibfnamefont {R.}~\bibnamefont
  {Blatt}},\ }\href {\doibase 10.1038/nature09801} {\bibfield  {journal}
  {\bibinfo  {journal} {Nature}\ }\textbf {\bibinfo {volume} {470}},\ \bibinfo
  {pages} {486} (\bibinfo {year} {2011})},\ \Eprint
  {http://arxiv.org/abs/1104.1146} {arXiv:1104.1146} \BibitemShut {NoStop}%
\bibitem [{\citenamefont {Murch}\ \emph {et~al.}(2013)\citenamefont {Murch},
  \citenamefont {Weber}, \citenamefont {Macklin},\ and\ \citenamefont
  {Siddiqi}}]{Murch2013}%
  \BibitemOpen
  \bibfield  {author} {\bibinfo {author} {\bibfnamefont {K.~W.}\ \bibnamefont
  {Murch}}, \bibinfo {author} {\bibfnamefont {S.~J.}\ \bibnamefont {Weber}},
  \bibinfo {author} {\bibfnamefont {C.}~\bibnamefont {Macklin}}, \ and\
  \bibinfo {author} {\bibfnamefont {I.}~\bibnamefont {Siddiqi}},\ }\href
  {\doibase 10.1038/nature12539} {\bibfield  {journal} {\bibinfo  {journal}
  {Nature}\ }\textbf {\bibinfo {volume} {502}},\ \bibinfo {pages} {211}
  (\bibinfo {year} {2013})},\ \Eprint {http://arxiv.org/abs/1305.7270}
  {arXiv:1305.7270} \BibitemShut {NoStop}%
\bibitem [{\citenamefont {Roch}\ \emph {et~al.}(2014)\citenamefont {Roch},
  \citenamefont {Schwartz}, \citenamefont {Motzoi}, \citenamefont {Macklin},
  \citenamefont {Vijay}, \citenamefont {Eddins}, \citenamefont {Korotkov},
  \citenamefont {Whaley}, \citenamefont {Sarovar},\ and\ \citenamefont
  {Siddiqi}}]{Roch2014}%
  \BibitemOpen
  \bibfield  {author} {\bibinfo {author} {\bibfnamefont {N.}~\bibnamefont
  {Roch}}, \bibinfo {author} {\bibfnamefont {M.~E.}\ \bibnamefont {Schwartz}},
  \bibinfo {author} {\bibfnamefont {F.}~\bibnamefont {Motzoi}}, \bibinfo
  {author} {\bibfnamefont {C.}~\bibnamefont {Macklin}}, \bibinfo {author}
  {\bibfnamefont {R.}~\bibnamefont {Vijay}}, \bibinfo {author} {\bibfnamefont
  {A.~W.}\ \bibnamefont {Eddins}}, \bibinfo {author} {\bibfnamefont {A.~N.}\
  \bibnamefont {Korotkov}}, \bibinfo {author} {\bibfnamefont {K.~B.}\
  \bibnamefont {Whaley}}, \bibinfo {author} {\bibfnamefont {M.}~\bibnamefont
  {Sarovar}}, \ and\ \bibinfo {author} {\bibfnamefont {I.}~\bibnamefont
  {Siddiqi}},\ }\href {\doibase 10.1103/PhysRevLett.112.170501} {\bibfield
  {journal} {\bibinfo  {journal} {Physical Review Letters}\ }\textbf {\bibinfo
  {volume} {112}},\ \bibinfo {pages} {170501} (\bibinfo {year} {2014})},\
  \Eprint {http://arxiv.org/abs/1402.1868} {arXiv:1402.1868} \BibitemShut
  {NoStop}%
\bibitem [{\citenamefont {Weber}\ \emph {et~al.}(2014)\citenamefont {Weber},
  \citenamefont {Chantasri}, \citenamefont {Dressel}, \citenamefont {Jordan},
  \citenamefont {Murch},\ and\ \citenamefont {Siddiqi}}]{Weber2014}%
  \BibitemOpen
  \bibfield  {author} {\bibinfo {author} {\bibfnamefont {S.~J.}\ \bibnamefont
  {Weber}}, \bibinfo {author} {\bibfnamefont {A.}~\bibnamefont {Chantasri}},
  \bibinfo {author} {\bibfnamefont {J.}~\bibnamefont {Dressel}}, \bibinfo
  {author} {\bibfnamefont {A.~N.}\ \bibnamefont {Jordan}}, \bibinfo {author}
  {\bibfnamefont {K.~W.}\ \bibnamefont {Murch}}, \ and\ \bibinfo {author}
  {\bibfnamefont {I.}~\bibnamefont {Siddiqi}},\ }\href {\doibase
  10.1038/nature13559} {\bibfield  {journal} {\bibinfo  {journal} {Nature}\
  }\textbf {\bibinfo {volume} {511}},\ \bibinfo {pages} {570} (\bibinfo {year}
  {2014})},\ \Eprint {http://arxiv.org/abs/1403.4992} {arXiv:1403.4992}
  \BibitemShut {NoStop}%
\bibitem [{\citenamefont {Balzani}\ \emph {et~al.}(2006)\citenamefont
  {Balzani}, \citenamefont {Credi}, \citenamefont {Silvi},\ and\ \citenamefont
  {Venturi}}]{Balzani2006}%
  \BibitemOpen
  \bibfield  {author} {\bibinfo {author} {\bibfnamefont {V.}~\bibnamefont
  {Balzani}}, \bibinfo {author} {\bibfnamefont {A.}~\bibnamefont {Credi}},
  \bibinfo {author} {\bibfnamefont {S.}~\bibnamefont {Silvi}}, \ and\ \bibinfo
  {author} {\bibfnamefont {M.}~\bibnamefont {Venturi}},\ }\href {\doibase
  10.1088/0953-8984/18/33/S01} {\bibfield  {journal} {\bibinfo  {journal}
  {Chemical Society reviews}\ }\textbf {\bibinfo {volume} {35}},\ \bibinfo
  {pages} {1135} (\bibinfo {year} {2006})}\BibitemShut {NoStop}%
\bibitem [{\citenamefont {Serreli}\ \emph {et~al.}(2007)\citenamefont
  {Serreli}, \citenamefont {Lee}, \citenamefont {Kay},\ and\ \citenamefont
  {Leigh}}]{Serreli2007}%
  \BibitemOpen
  \bibfield  {author} {\bibinfo {author} {\bibfnamefont {V.}~\bibnamefont
  {Serreli}}, \bibinfo {author} {\bibfnamefont {C.-F.}\ \bibnamefont {Lee}},
  \bibinfo {author} {\bibfnamefont {E.~R.}\ \bibnamefont {Kay}}, \ and\
  \bibinfo {author} {\bibfnamefont {D.~A.}\ \bibnamefont {Leigh}},\ }\href
  {\doibase 10.1038/nature05452} {\bibfield  {journal} {\bibinfo  {journal}
  {Nature}\ }\textbf {\bibinfo {volume} {445}},\ \bibinfo {pages} {523}
  (\bibinfo {year} {2007})}\BibitemShut {NoStop}%
\bibitem [{\citenamefont {Yildiz}\ \emph {et~al.}(2008)\citenamefont {Yildiz},
  \citenamefont {Tomishige}, \citenamefont {Gennerich},\ and\ \citenamefont
  {Vale}}]{Yildiz2008}%
  \BibitemOpen
  \bibfield  {author} {\bibinfo {author} {\bibfnamefont {A.}~\bibnamefont
  {Yildiz}}, \bibinfo {author} {\bibfnamefont {M.}~\bibnamefont {Tomishige}},
  \bibinfo {author} {\bibfnamefont {A.}~\bibnamefont {Gennerich}}, \ and\
  \bibinfo {author} {\bibfnamefont {R.~D.}\ \bibnamefont {Vale}},\ }\href
  {\doibase 10.1016/j.cell.2008.07.018} {\bibfield  {journal} {\bibinfo
  {journal} {Cell}\ }\textbf {\bibinfo {volume} {134}},\ \bibinfo {pages}
  {1030} (\bibinfo {year} {2008})}\BibitemShut {NoStop}%
\bibitem [{\citenamefont {Toyabe}\ \emph {et~al.}(2012)\citenamefont {Toyabe},
  \citenamefont {Ueno},\ and\ \citenamefont {Muneyuki}}]{Toyabe2012}%
  \BibitemOpen
  \bibfield  {author} {\bibinfo {author} {\bibfnamefont {S.}~\bibnamefont
  {Toyabe}}, \bibinfo {author} {\bibfnamefont {H.}~\bibnamefont {Ueno}}, \ and\
  \bibinfo {author} {\bibfnamefont {E.}~\bibnamefont {Muneyuki}},\ }\href
  {\doibase 10.1209/0295-5075/97/40004} {\bibfield  {journal} {\bibinfo
  {journal} {EPL (Europhysics Letters)}\ }\textbf {\bibinfo {volume} {97}},\
  \bibinfo {pages} {40004} (\bibinfo {year} {2012})},\ \Eprint
  {http://arxiv.org/abs/arXiv:1112.0186v1} {arXiv:arXiv:1112.0186v1}
  \BibitemShut {NoStop}%
\bibitem [{\citenamefont {Lan}\ \emph {et~al.}(2012)\citenamefont {Lan},
  \citenamefont {Sartori}, \citenamefont {Neumann}, \citenamefont {Sourjik},\
  and\ \citenamefont {Tu}}]{Lan2012}%
  \BibitemOpen
  \bibfield  {author} {\bibinfo {author} {\bibfnamefont {G.}~\bibnamefont
  {Lan}}, \bibinfo {author} {\bibfnamefont {P.}~\bibnamefont {Sartori}},
  \bibinfo {author} {\bibfnamefont {S.}~\bibnamefont {Neumann}}, \bibinfo
  {author} {\bibfnamefont {V.}~\bibnamefont {Sourjik}}, \ and\ \bibinfo
  {author} {\bibfnamefont {Y.}~\bibnamefont {Tu}},\ }\href {\doibase
  10.1038/nphys2276} {\bibfield  {journal} {\bibinfo  {journal} {Nature
  Physics}\ }\textbf {\bibinfo {volume} {8}},\ \bibinfo {pages} {422} (\bibinfo
  {year} {2012})},\ \Eprint {http://arxiv.org/abs/0402594v3} {arXiv:0402594v3
  [arXiv:cond-mat]} \BibitemShut {NoStop}%
\bibitem [{\citenamefont {Kawai}\ \emph {et~al.}(2007)\citenamefont {Kawai},
  \citenamefont {Parrondo},\ and\ \citenamefont {den Broeck}}]{Kawai2007}%
  \BibitemOpen
  \bibfield  {author} {\bibinfo {author} {\bibfnamefont {R.}~\bibnamefont
  {Kawai}}, \bibinfo {author} {\bibfnamefont {J.~M.~R.}\ \bibnamefont
  {Parrondo}}, \ and\ \bibinfo {author} {\bibfnamefont {C.~V.}\ \bibnamefont
  {den Broeck}},\ }\href {\doibase 10.1103/PhysRevLett.98.080602} {\bibfield
  {journal} {\bibinfo  {journal} {Physical Review Letters}\ }\textbf {\bibinfo
  {volume} {98}},\ \bibinfo {pages} {080602} (\bibinfo {year} {2007})},\
  \Eprint {http://arxiv.org/abs/0701397} {arXiv:0701397 [cond-mat]}
  \BibitemShut {NoStop}%
\bibitem [{\citenamefont {Tian}(2009)}]{Tian2009}%
  \BibitemOpen
  \bibfield  {author} {\bibinfo {author} {\bibfnamefont {L.}~\bibnamefont
  {Tian}},\ }\href {\doibase 10.1103/PhysRevB.79.193407} {\bibfield  {journal}
  {\bibinfo  {journal} {Physical Review B - Condensed Matter and Materials
  Physics}\ }\textbf {\bibinfo {volume} {79}},\ \bibinfo {pages} {1} (\bibinfo
  {year} {2009})},\ \Eprint {http://arxiv.org/abs/0809.4459} {arXiv:0809.4459}
  \BibitemShut {NoStop}%
\bibitem [{\citenamefont {Kutvonen}\ \emph {et~al.}(2016)\citenamefont
  {Kutvonen}, \citenamefont {Sagawa},\ and\ \citenamefont
  {Ala-Nissila}}]{Kutvonen2015}%
  \BibitemOpen
  \bibfield  {author} {\bibinfo {author} {\bibfnamefont {A.}~\bibnamefont
  {Kutvonen}}, \bibinfo {author} {\bibfnamefont {T.}~\bibnamefont {Sagawa}}, \
  and\ \bibinfo {author} {\bibfnamefont {T.}~\bibnamefont {Ala-Nissila}},\
  }\href {\doibase 10.1103/PhysRevE.93.032147} {\bibfield  {journal} {\bibinfo
  {journal} {Physical Review E}\ }\textbf {\bibinfo {volume} {93}},\ \bibinfo
  {pages} {032147} (\bibinfo {year} {2016})},\ \Eprint
  {http://arxiv.org/abs/1510.00190} {arXiv:1510.00190} \BibitemShut {NoStop}%
\bibitem [{\citenamefont {Nakatani}\ and\ \citenamefont
  {Ogawa}(2010)}]{Nakatani2010}%
  \BibitemOpen
  \bibfield  {author} {\bibinfo {author} {\bibfnamefont {M.}~\bibnamefont
  {Nakatani}}\ and\ \bibinfo {author} {\bibfnamefont {T.}~\bibnamefont
  {Ogawa}},\ }\href {\doibase 10.1143/JPSJ.79.084401} {\bibfield  {journal}
  {\bibinfo  {journal} {Journal of the Physical Society of Japan}\ }\textbf
  {\bibinfo {volume} {79}},\ \bibinfo {pages} {084401} (\bibinfo {year}
  {2010})}\BibitemShut {NoStop}%
\bibitem [{\citenamefont {Bartolotta}\ \emph {et~al.}(2016)\citenamefont
  {Bartolotta}, \citenamefont {Carroll}, \citenamefont {Leichenauer},\ and\
  \citenamefont {Pollack}}]{Bartolotta2015}%
  \BibitemOpen
  \bibfield  {author} {\bibinfo {author} {\bibfnamefont {A.}~\bibnamefont
  {Bartolotta}}, \bibinfo {author} {\bibfnamefont {S.~M.}\ \bibnamefont
  {Carroll}}, \bibinfo {author} {\bibfnamefont {S.}~\bibnamefont
  {Leichenauer}}, \ and\ \bibinfo {author} {\bibfnamefont {J.}~\bibnamefont
  {Pollack}},\ }\href {\doibase 10.1103/PhysRevE.94.022102} {\bibfield
  {journal} {\bibinfo  {journal} {Physical Review E}\ }\textbf {\bibinfo
  {volume} {94}},\ \bibinfo {pages} {022102} (\bibinfo {year} {2016})},\
  \Eprint {http://arxiv.org/abs/1508.02421} {arXiv:1508.02421} \BibitemShut
  {NoStop}%
\bibitem [{\citenamefont {Esposito}(2012)}]{Esposito2012}%
  \BibitemOpen
  \bibfield  {author} {\bibinfo {author} {\bibfnamefont {M.}~\bibnamefont
  {Esposito}},\ }\href {\doibase 10.1103/PhysRevE.85.041125} {\bibfield
  {journal} {\bibinfo  {journal} {Physical Review E - Statistical, Nonlinear,
  and Soft Matter Physics}\ }\textbf {\bibinfo {volume} {85}},\ \bibinfo
  {pages} {1} (\bibinfo {year} {2012})},\ \Eprint
  {http://arxiv.org/abs/1112.5410} {arXiv:1112.5410} \BibitemShut {NoStop}%
\bibitem [{\citenamefont {Mehl}\ \emph {et~al.}(2012)\citenamefont {Mehl},
  \citenamefont {Lander}, \citenamefont {Bechinger}, \citenamefont {Blickle},\
  and\ \citenamefont {Seifert}}]{Mehl2012}%
  \BibitemOpen
  \bibfield  {author} {\bibinfo {author} {\bibfnamefont {J.}~\bibnamefont
  {Mehl}}, \bibinfo {author} {\bibfnamefont {B.}~\bibnamefont {Lander}},
  \bibinfo {author} {\bibfnamefont {C.}~\bibnamefont {Bechinger}}, \bibinfo
  {author} {\bibfnamefont {V.}~\bibnamefont {Blickle}}, \ and\ \bibinfo
  {author} {\bibfnamefont {U.}~\bibnamefont {Seifert}},\ }\href {\doibase
  10.1103/PhysRevLett.108.220601} {\bibfield  {journal} {\bibinfo  {journal}
  {Physical Review Letters}\ }\textbf {\bibinfo {volume} {108}},\ \bibinfo
  {pages} {1} (\bibinfo {year} {2012})},\ \Eprint
  {http://arxiv.org/abs/1205.0238} {arXiv:1205.0238} \BibitemShut {NoStop}%
\bibitem [{\citenamefont {Kawaguchi}\ and\ \citenamefont
  {Nakayama}(2013)}]{Kawaguchi2013}%
  \BibitemOpen
  \bibfield  {author} {\bibinfo {author} {\bibfnamefont {K.}~\bibnamefont
  {Kawaguchi}}\ and\ \bibinfo {author} {\bibfnamefont {Y.}~\bibnamefont
  {Nakayama}},\ }\href {\doibase 10.1103/PhysRevE.88.022147} {\bibfield
  {journal} {\bibinfo  {journal} {Physical Review E}\ }\textbf {\bibinfo
  {volume} {88}},\ \bibinfo {pages} {022147} (\bibinfo {year} {2013})},\
  \Eprint {http://arxiv.org/abs/1209.6333} {arXiv:1209.6333} \BibitemShut
  {NoStop}%
\bibitem [{\citenamefont {Suzuki}(1977)}]{Suzuki1977}%
  \BibitemOpen
  \bibfield  {author} {\bibinfo {author} {\bibfnamefont {M.}~\bibnamefont
  {Suzuki}},\ }\href {\doibase 10.1007/BF01614161} {\bibfield  {journal}
  {\bibinfo  {journal} {Communications in Mathematical Physics}\ }\textbf
  {\bibinfo {volume} {57}},\ \bibinfo {pages} {193} (\bibinfo {year}
  {1977})}\BibitemShut {NoStop}%
\bibitem [{\citenamefont {Casas}\ \emph {et~al.}(2012)\citenamefont {Casas},
  \citenamefont {Murua},\ and\ \citenamefont {Nadinic}}]{Casas2012}%
  \BibitemOpen
  \bibfield  {author} {\bibinfo {author} {\bibfnamefont {F.}~\bibnamefont
  {Casas}}, \bibinfo {author} {\bibfnamefont {A.}~\bibnamefont {Murua}}, \ and\
  \bibinfo {author} {\bibfnamefont {M.}~\bibnamefont {Nadinic}},\ }\href
  {\doibase 10.1016/j.cpc.2012.06.006} {\bibfield  {journal} {\bibinfo
  {journal} {Computer Physics Communications}\ }\textbf {\bibinfo {volume}
  {183}},\ \bibinfo {pages} {2386} (\bibinfo {year} {2012})},\ \Eprint
  {http://arxiv.org/abs/arXiv:1204.0389v2} {arXiv:arXiv:1204.0389v2}
  \BibitemShut {NoStop}%
\end{thebibliography}%

\onecolumngrid
\appendix

\section{Time-reversal in the modified QJ approach\label{app:TimeReversal}}
Let us consider the time reversal of the reverse dynamics in the modified QJ approach.  We define the time-reversal superoperator in Liouville space $\vartheta:= \Theta^T\otimes\Theta$, such that the time reversal of an operator in Hilbert space $\Theta\rho\Theta^{-1}$ becomes $\vartheta\kket{\rho}$ in Liouville space, and a superoperator of the form $A(\cdot)B$ for some operators $A,B$, is time-reversed as $\vartheta (B^T \otimes A) \vartheta^{-1} vec(\cdot)$ in Liouville space.

The generator of the evolution between visible transitions is 
\begin{equation}
\G = \H + \L_{{\rm nj}} + \L_Y
\end{equation}
with
\begin{eqnarray}
\H &=& -i(\I\otimes H - H^T\otimes\I) \\
\L_{{\rm nj}} &=& -\frac{1}{2}\sum_k\Bigl(\I\otimes L_k^{\dagger}L_k + L_k^{T}L_k^*\otimes\I\Bigr) \\
\L_Y &=& \sum_{k\in k_Y} L_{k}^* \otimes L_k.
\end{eqnarray}
Due to the anti-unitarity of $\Theta$ we do not simply have $\tilde{\G} = \vartheta \G \vartheta^{-1}$. Instead, due to the $i$ in the definition of $\H$ we have
\begin{eqnarray}
\tilde{\G} &=& \vartheta \Bigl( -\H + \L_{{\rm nj}} + \L_Y \Bigr) \vartheta^{-1} \\
&=:& \vartheta\bar{\G} \vartheta^{-1}.
\end{eqnarray}
This $\tilde{\G}$ is the Liouville space generator of evolution between visible jumps for the time reversed process, such that  \mbox{$\kket{\tilde{\rho}(t+\Delta t)} = e^{\tilde{\G} \Delta t} \kket{\tilde{\rho}(t)}$} if there was no visible jump between times $t$ and $t+\Delta t$. As we have done in the main text for the forward trajectory, we can now construct the state at the end of the trajectory just before the final measurement for the reverse trajectory $\tilde{\tau}_X$, finding
\begin{equation}
\kket{\tilde{\rho}(T)} = e^{\tilde{\G} \Delta t_0}\tilde{\J}_{\tilde{k}_{1}} ... \tilde{\J}_{\tilde{k}_{M}}e^{\tilde{\G} \Delta t_M}\kket{\tilde{b}}dt^M.
\end{equation}
Thus after performing the final measurement and finding the state $\ket{a}$ we have for the probability of the reverse trajectory
\begin{eqnarray}
P(\tilde{\tau}_X | \tilde{b}) &=& \bbra{\tilde{a}} e^{\tilde{\G} \Delta t_0}\tilde{\J}_{\tilde{k}_{1}} ... \tilde{\J}_{\tilde{k}_{M}}e^{\tilde{\G} \Delta t_M}\kket{\tilde{b}}dt^M\\
&=& \bbra{a} \vartheta^{-1}\vartheta e^{\bar{\G} \Delta t_0}\vartheta^{-1}\vartheta\J_{\tilde{k}_{1}}\vartheta^{-1} ... \vartheta\J_{\tilde{k}_{M}}\vartheta^{-1}\vartheta e^{\bar{\G} \Delta t_M}\vartheta^{-1}\vartheta\kket{b}dt^M\\
&=& \bbra{a} e^{\bar{\G} \Delta t_0}\J_{\tilde{k}_{1}} ... \J_{\tilde{k}_{M}} e^{\bar{\G} \Delta t_M}\kket{b}dt^M
\end{eqnarray}
Now since the probability $P(\tilde{\tau}_X | \tilde{b}) \in \mathbb{R}$ we can take the complex conjugate to get
\begin{eqnarray}\label{eq:probAlmostDone}
P(\tilde{\tau}_X | \tilde{b}) &=& \bbra{b}  e^{\mathcal{\bar{G}^{\dagger}}\Delta t_{M}} \mathcal{J}^{\dagger}_{\tilde{k}_{M}}...\mathcal{J}^{\dagger}_{\tilde{k}_1} e^{\mathcal{\bar{G}^{\dagger}}\Delta t_0}\kket{a} dt^{M}
\end{eqnarray}
Reminding ourselves of the definition of the visible jump superoperators $\J_k = L_{k}^* \otimes L_{k}$ we see
\begin{eqnarray}
\J_{\tilde{k}}^{\dagger} &=& (L_{\tilde{k}}^{\dagger})^* \otimes L_{\tilde{k}}^{\dagger} \\
&=& L_{k}^* \otimes L_{k} e^{-\Delta s_k}\\
&=& \J_k e^{-\Delta s_k}
\end{eqnarray}
where in the second line we have used the operator detailed balance condition Eq. (\ref{eq:DetailedBalance}). Substituting into (\ref{eq:probAlmostDone}) we arrive at
\begin{eqnarray}
P(\tilde{\tau}_X | \tilde{b}) &=& \bbra{b}  e^{\mathcal{\bar{G}^{\dagger}}\Delta t_{M}} \mathcal{J}_{k_{M}}...\mathcal{J}_{k_1} e^{\mathcal{\bar{G}^{\dagger}}\Delta t_0}\kket{a} dt^{M} e^{-\Delta s_{{\rm env}}(\tau_X)}
\end{eqnarray}
which concludes the derivation of Eq. (\ref{eq:PVisibleBackward}). The operator $\bar{\G}^{\dagger}$ appearing in this expression as well as the fundamental expression for the coarse-grained hidden entropy Eq. (\ref{eq:DeltaIY}) can easily be shown to equal
\begin{equation}
\bar{\G}^{\dagger} = \H + \L_{{\rm nj}} + \L_Y^{\dagger},
\end{equation}
since $\H^{\dagger} = -\H$, $\L_{{\rm nj}}^{\dagger} = \L_{{\rm nj}}$, but in general $\L_Y^{\dagger} \neq \L_Y$,  i.e. it differs from $\G$ only in the term related to the hidden jumps. This also proves that if all jumps are visible, i.e. $k_Y = \emptyset$ and hence $\L_Y$ vanishes, the hidden entropy production also vanishes since $P(\tilde{\tau}_X | \tilde{b}) = P(\tau_X | a) e^{-\Delta s_{{\rm env}}(\tau_X)}$ and the ratio in (\ref{eq:DeltaIY}) cancels. We then simply recover the already well known results reviewed in section \ref{sec:QJ} for fully visible QJ trajectories.

\section{Explicit derivations for the two qubit Maxwell's Demon\label{app:ExampleDerivation}}
In this section we provide the explicit derivations for the model presented in section \ref{sec:Demon}, particularly the expression (\ref{DeltaID_medium}) for ${\Delta \sigma}_Y$. The approach will be to separate the exponentials \mbox{$\exp[\G\Delta t_i] = \exp[(\H_{{\rm eff}} + \L_Y)\Delta t_i]$} and \mbox{$\exp[\bar{\G}^{\dagger}\Delta t_i] = \exp[(\H_{{\rm eff}} + \L_Y^{\dagger})\Delta t_i]$}  appearing in the Eq. (\ref{eq:DeltaIY}), where we have defined $\H_{{\rm eff}} := \H + \L_{{\rm nj}}$ as the Liouville space equivalent of the effective Hamiltonian $H_{{\rm eff}}$. Our main tool for achieving this is the Zassenhaus formula \cite{Suzuki1977}
\begin{equation}\label{Zassenhaus}
e^{t(A+B)} = e^{t A} e^{t B} e^{-\frac{t^2}{2!} [A,B]}e^{\frac{t^3}{3!} (2[B,[A,B]] + [A,[A,B]])} ...
\end{equation}
for matrices $A$ and $B$, where in our case $A = \mathcal{H}_{{\rm eff}}$, and $B= \mathcal{L_Y}$ or $B= \mathcal{L_Y}^{\dagger}$. Note that in the following we will only consider the case $\lambda = 0$.

Let us first introduce a result relating to the commutators of the jump operators which we will employ later. As noted in the main text, the demon operators do not connect unique states if $\gamma_Y > 0$. However, we can split the demon operators into two parts
\begin{equation}
L_{k_y} =: L_{k_y}^e + L_{k_y}^g
\end{equation}
where $L_{k_y}^{e/g}$ refers to the term in $L_{k_y}$ that is controlled by the system's $\ket{e/g}$ state. As an example, for \mbox{$L_5 = \gamma_5 (\ket{e}\bra{e} + \gamma_Y \ket{g}\bra{g})\otimes \sigma_-$} we have \mbox{$L_5^e= \gamma_5 \ket{e}\bra{e}\otimes \sigma_-$} and \mbox{$L_5^g= \gamma_5 \gamma_Y \ket{g}\bra{g}\otimes \sigma_-$}. It can be shown that
\begin{equation} \label{eq:commutatorSplit}
\sum_{k}  [L_k^{\dagger} L_k, L_{k_y}^{e/g}] = \alpha_{k_y}^{e/g} L_{k_y}^{e/g},
\end{equation}
where the sum over $k$ includes both visible and hidden transitions, where
\begin{eqnarray}
\alpha_5^e &=& (1-\gamma_X^2)(\gamma_3^2 - \gamma_1^2) + (\gamma_6^2 - \gamma_5^2) + \gamma_Y^2 (\gamma_8^2 - \gamma_7^2) \\
\alpha_5^g &=& (1-\gamma_X^2)(\gamma_4^2 - \gamma_2^2) + (\gamma_8^2 - \gamma_7^2) + \gamma_Y^2 (\gamma_6^2 - \gamma_5^2) 
\end{eqnarray}
and 
\begin{equation}
\alpha_5^x = - \alpha_6^x = \alpha_7^x = - \alpha_8^x
\end{equation}
for $x\in\{e,g\}$. It straightforwardly follows that
\begin{equation}
\sum_{k}  [L_k^{\dagger} L_k, L_{k_y}] = \sum_{x\in\{e,g\}}\alpha_{k_y}^{x} L_{k_y}^{x}.
\end{equation}
We can further show that (if $\lambda = 0$) 
\begin{equation}
[\H,\L_Y] = 0 = [\H,\L_Y^{\dagger}]
\end{equation}
The first commutator we have to evaluate $[\H+\L_{{\rm nj}},\L_Y]$ thus simplifies to $[\L_{{\rm nj}},\L_Y]$. Using this and Eq. (\ref{eq:commutatorSplit}) we can show that
\begin{eqnarray}
[\mathcal{H}_{{\rm eff}},\mathcal{L}_Y]_{(n)} = (-1)^n \sum_{k\in k_Y} \sum_{x,x'\in\{e,g\}} \Bigl(\frac{\alpha_{k}^x + \alpha_{k}^{x'}}{2}\Bigr)^n L_{k}^{x} \otimes L_{k}^{x'}
\end{eqnarray}
and
\begin{eqnarray}
[\mathcal{H}_{{\rm eff}},\mathcal{L}_Y^{\dagger}]_{(n)} = (-1)^n \sum_{k\in k_Y} \sum_{x,x'\in\{e,g\}} \Bigl(\frac{\alpha_{k}^x + \alpha_{k}^{x'}}{2}\Bigr)^n L_{k}^{x} \otimes L_{k}^{x'} e^{-\Delta s_{k_D}} 
\end{eqnarray}
where we have used $L_{k}^* = L_{k}$ for all jump operators, and introduced the notation
\begin{equation}
[A,B]_{(n)} := \left[A, \left[A, \left[A,...,\left[A,B\right]\right]\right]\right]
\end{equation}
for nested commutators such that $[A,B]_{(0)} =B$, $[A,B]_{(1)} = [A,B]$, $[A,B]_{(2)} = [A,[A,B]]$, etc. The only other non-vanishing commutators are the nested commutators $[\L_Y, [\mathcal{H}_{{\rm eff}},\mathcal{L}_Y^{\dagger}]_{(n)}]$. Let us first introduce a shorthand for the diagonal elements in Liouville space,
\begin{eqnarray}
\Box_{x,x'}^{y,y'} := \ket{x,y}\bra{x,y} \otimes \ket{x',y'}\bra{x'y'} \qquad \text{for} \qquad x,x'\in\{e,g\}; y,y'\in\{0,1\}
\end{eqnarray}
such that
\begin{eqnarray}
\mathbb{1} = \sum_{x,x',y,y'} \Box_{x,x'}^{y,y'}.
\end{eqnarray}
Using this notation we find
\begin{eqnarray}
[\mathcal{L}_Y,[\mathcal{H}_{{\rm eff}},\mathcal{L}_Y]_{(n)}] &=& (-1)^n \sum_{k,k'\in k_Y} \sum_{x,{x'},{x^*},x^+\in\{e,g\}} \Bigl(\frac{\alpha_{k}^x + \alpha_{k}^{x'}}{2}\Bigr)^n [L_{{k'}}^{x} \otimes L_{{k'}}^{x'},L_{k}^{{x^*}} \otimes L_{k}^{x^+}]  \\
&=& (-1)^n \sum_{k,k'\in k_Y} \sum_{x,{x'},{x^*},x^+\in\{e,g\}} \Bigl(\frac{\alpha_{k}^x + \alpha_{k}^{x'}}{2}\Bigr)^n [L_{{k'}}^{x} \otimes L_{{k'}}^{x'},L_{k}^{{x^*}} \otimes L_{k}^{x^+}]\delta_{x{x^*}}\delta_{x'x^+}  \\
&=& (-1)^n \sum_{k,k'\in k_Y} \sum_{x,{x'}\in\{e,g\}} \Bigl(\frac{\alpha_{k}^x + \alpha_{k}^{x'}}{2}\Bigr)^n [L_{{k'}}^{x} \otimes L_{{k'}}^{x'},L_{k}^{x} \otimes L_{k}^{{x'}}]  \\
&=& (-2\alpha_5^e)^n  \Bigl( \gamma_5^2\gamma_6^2 +  \gamma_Y^2(\gamma_5^2\gamma_8^2 +\gamma_6^2\gamma_7^2) + \gamma_Y^4\gamma_7^2\gamma_8^2 \Bigr)(\Box_{ee}^{11}-\Box_{ee}^{00})\nonumber\\
&&+ (-2\alpha_5^g)^n  \Bigl( \gamma_Y^4\gamma_5^2\gamma_6^2 +  \gamma_Y^2(\gamma_5^2\gamma_8^2 +\gamma_6^2\gamma_7^2) + \gamma_7^2\gamma_8^2 \Bigr)(\Box_{gg}^{11}-\Box_{gg}^{00})\nonumber\\
&&+ ((-1)^n + 1) \frac{\alpha_{5}^e + \alpha_{5}^g}{2} \gamma_Y^2 \Bigl(\gamma_5^2\gamma_6^2 +  \gamma_5^2\gamma_8^2 +\gamma_6^2\gamma_7^2 + \gamma_7^2\gamma_8^2 \Bigr)(\Box_{eg}^{11}-\Box_{eg}^{00} + \Box_{ge}^{11}-\Box_{ge}^{00}), \label{eq:oddComm}
\end{eqnarray}
and similarly for the backward case
\begin{eqnarray}
[\mathcal{L}_Y^{\dagger},[\mathcal{H}_{{\rm eff}},\mathcal{L}_Y^{\dagger}]_{(n)}] = (-2\alpha_5^e)^n  \Bigl( \gamma_5^2\gamma_6^2 +  \gamma_Y^2(\gamma_5^2\gamma_8^2 e^{-(\Delta s_5 + \Delta s_8)} +\gamma_6^2\gamma_7^2e^{+(\Delta s_5 + \Delta s_8)}) + \gamma_Y^4\gamma_7^2\gamma_8^2 \Bigr)(\Box_{ee}^{11}-\Box_{ee}^{00})\nonumber\\
+ (-2\alpha_5^g)^n  \Bigl( \gamma_Y^4\gamma_5^2\gamma_6^2 +  \gamma_Y^2(\gamma_5^2\gamma_8^2e^{-(\Delta s_5 + \Delta s_8)} +\gamma_6^2\gamma_7^2e^{+(\Delta s_5 + \Delta s_8)}) + \gamma_7^2\gamma_8^2 \Bigr)(\Box_{gg}^{11}-\Box_{gg}^{00})\nonumber\\
+ ((-1)^n + 1) \frac{\alpha_{5}^e + \alpha_{5}^g}{2} \gamma_Y^2 \Bigl(\gamma_5^2\gamma_6^2 +  \gamma_5^2\gamma_8^2e^{-(\Delta s_5 + \Delta s_8)} +\gamma_6^2\gamma_7^2e^{+(\Delta s_5 + \Delta s_8)} + \gamma_7^2\gamma_8^2 \Bigr)(\Box_{eg}^{11}-\Box_{eg}^{00} + \Box_{ge}^{11}-\Box_{ge}^{00}). \label{eq:oddCommBack}
\end{eqnarray}
Noting that $\gamma_5^2\gamma_8^2 e^{-(\Delta s_5 + \Delta s_8)} = \gamma_6^2\gamma_7^2$ we see that this is actually the same as in the forward process, i.e. $[\mathcal{L}_Y^{\dagger},[\mathcal{H}_{{\rm eff}},\mathcal{L}_Y^{\dagger}]_{(n)}] = [\mathcal{L}_Y,[\mathcal{H}_{{\rm eff}},\mathcal{L}_Y]_{(n)}]$. Despite looking rather unappealing, these commutators have the nice properties of being diagonal and commuting with all the other operators of interest. Particularly 
\begin{eqnarray}
[[\mathcal{H}_{{\rm eff}},\mathcal{L}_Y]_{(n)},[\mathcal{L}_Y,[\mathcal{H}_{{\rm eff}},\mathcal{L}_Y]_{(m)}]] = 0 \quad \forall \quad n,m  \geq 1.
\end{eqnarray}
Considering again the Zassenhaus formula 
\begin{eqnarray}
e^{t(\mathcal{H}_{{\rm eff}}+\mathcal{L}_Y)} = e^{t \mathcal{H}_{{\rm eff}}} e^{t \mathcal{L}_Y} e^{-\frac{t^2}{2!} [\mathcal{H}_{{\rm eff}},\mathcal{L}_Y]}e^{\frac{t^3}{3!} (2[\mathcal{L}_Y,[\mathcal{H}_{{\rm eff}},\mathcal{L}_Y]] + [\mathcal{H}_{{\rm eff}},[\mathcal{H}_{{\rm eff}},\mathcal{L}_Y]])} ... ,
\end{eqnarray}
these vanishing commutators allow us to split the exponentials at the $n^{th}$ order into two factors
\begin{eqnarray}
e^{a_n [\mathcal{H}_{{\rm eff}},\mathcal{L}_Y]_{(n)} + b_n {[}\mathcal{L}_Y,[\mathcal{H}_{{\rm eff}},\mathcal{L}_Y]_{(n-1)}{]}} &=& e^{a_n [\mathcal{H}_{{\rm eff}},\mathcal{L}_Y]_{(n)}}e^{b_n {[}\mathcal{L}_Y,[\mathcal{H}_{{\rm eff}},\mathcal{L}_Y]_{(n-1)}{]}}
\end{eqnarray}
and further collect all $a_n$ and all $b_n$ factors together such that
\begin{eqnarray}\label{eq:ZassenhausGen1}
e^{t(\mathcal{H}_{{\rm eff}}+\mathcal{L}_Y)} = e^{t {H}_{{\rm eff}}}\prod_{n=0}^{\infty}e^{a_n \frac{t^{n+1}}{(n+1)!}[\mathcal{H}_{{\rm eff}},\mathcal{L}_Y]_{(n)}}\prod_{m=1}^{\infty}e^{b_{m} \frac{t^{m+1}}{(m+1)!} {[}\mathcal{L}_Y,[\mathcal{H}_{{\rm eff}},\mathcal{L}_Y]_{(m-1)}{]}}.
\end{eqnarray}
In general, computing the coefficients in the Zassenhaus formula is a non-trivial problem that can only be solved procedurally \cite{Suzuki1977,Casas2012}. However, the $a_n$ coefficients of the 'most nested' commutators are special in that they take the simple form $a_n = (-1)^n$. Also, even though we do not know the explicit form of the $b_n$ coefficients we know that due to the diagonal form and commutative nature of the commutators in Eq. (\ref{eq:oddComm}) that the factor
\begin{eqnarray}
\xi := \prod_{m=1}^{\infty}e^{b_{m} \frac{t^{m+1}}{(m+1)!} {[}\mathcal{L}_Y,[\mathcal{H}_{{\rm eff}},\mathcal{L}_Y]_{(m-1)}{]}}
\end{eqnarray}
also has the same nice properties. The commutators $[[\mathcal{H}_{{\rm eff}},\mathcal{L}_Y]_{(n)},[\mathcal{H}_{{\rm eff}},\mathcal{L}_Y]_{(m)}]$ have the same diagonal form as Eq. (\ref{eq:oddComm}), only with slightly different terms on the diagonal. This allows us to turn the product of exponentials in Eq. (\ref{eq:ZassenhausGen1}) into an exponential sum times some additional diagonal and commuting factor. We get
\begin{eqnarray}\label{eq:ZassenhausGen2}
e^{t(\mathcal{H}_{{\rm eff}}+\mathcal{L}_Y)} = e^{t {H}_{{\rm eff}}}e^{\sum_{n=0}^{\infty} (-1)^n \frac{t^{n+1}}{(n+1)!}[\mathcal{H}_{{\rm eff}},\mathcal{L}_Y]_{(n)}} \xi'
\end{eqnarray}
where $\xi'$ collects all the diagonal and commuting factors and is of the form $\xi' = \sum_{x,x',y,y'} c_{xx'}^{yy'} \Box _{xx'}^{yy'}$ for some coefficients $c_{xx'}^{yy'}$. Substituting the explicit expression for the commutators and computing the sum over $n$ this reads
\begin{eqnarray}\label{eq:ZassenhausGen3}
e^{t(\mathcal{H}_{{\rm eff}}+\mathcal{L}_Y)} = e^{t {H}_{{\rm eff}}} \xi' \exp\Bigl( \sum_{k\in k_Y} \sum_{x,{x'}\in\{e,g\}} 2\frac{e^{\frac{\alpha_{k}^x + \alpha_{k}^{x'}}{2}t}-1}{\alpha_{k}^x + \alpha_{k}^{x'}} L_{k}^x \otimes L_{k}^{x'}\Bigr).
\end{eqnarray}
In the following we will refer to the exponential term as $B(t)$ which we further simplify by introducing the shorthand notation
\begin{eqnarray}
A_{k}^{xx'}(t) := 2\frac{e^{\frac{\alpha_{k}^x + \alpha_{k}^{x'}}{2}t}-1}{\alpha_{k}^x + \alpha_{k}^{x'}}
\end{eqnarray}
such that
\begin{eqnarray}
B(t) = \exp\Bigl( \sum_{k\in k_Y} \sum_{x,{x'}\in\{e,g\}} A_{k}^{x{x'}}(t) L_{k}^x \otimes L_{k}^{x'}\Bigr).
\end{eqnarray}
Note that due to the relation of the $\alpha$ parameters we have $A_{5}^{xx'}(t) = A_{7}^{xx'}(t)$ and $A_{6}^{xx'}(t) = A_{8}^{xx'}(t)$. Let us further define
\begin{eqnarray}
C_{kk'}^{xx'}(t) = \gamma_{k}^2\gamma_{k'}^2 A_{k}^{xx'}(t)A_{k'}^{xx'}(t)
\end{eqnarray}
and introduce one final shorthand notation, $\mathcal{A}_{xx'}(t)$, where
\begin{eqnarray}
\mathcal{A}_{ee}(t) = C_{56}^{ee}(t) + \gamma_Y^2(C_{58}^{ee}(t) + C_{67}^{ee}(t)) + \gamma_Y^4C_{78}^{ee}(t) \\
\mathcal{A}_{gg}(t) = \gamma_Y^4 C_{56}^{gg}(t) + \gamma_Y^2(C_{58}^{gg}(t) + C_{67}^{gg}(t)) + C_{78}^{gg}(t) \\
\mathcal{A}_{ge}(t) = \gamma_Y^2\Bigl(C_{56}^{ge}(t) + C_{58}^{ge}(t) + C_{67}^{ge}(t) + \gamma_Y^4C_{78}^{ge}(t)\Bigr) \\
\mathcal{A}_{eg}(t) = \gamma_Y^2\Bigl(C_{56}^{eg}(t) + C_{58}^{eg}(t) + C_{67}^{eg}(t) + \gamma_Y^4C_{78}^{eg}(t)\Bigr).
\end{eqnarray}
Finally, expanding the exponential $B(t)$ we arrive at the simple expression
\begin{eqnarray}
B(t) = \sum_{x,x'}\Bigl(\Box_{x{x'}}^{01} + \Box_{x{x'}}^{10} + \cosh\sqrt{\mathcal{A}_{x{x'}}(t)} (\Box_{x{x'}}^{00} + \Box_{x{x'}}^{11})\Bigr) + \sum_{x,{x'}}  \frac{\sinh\sqrt{\mathcal{A}_{x{x'}}(t)}}{\sqrt{\mathcal{A}_{x{x'}}(t)}} \sum_{k \in k_Y}A_{k}^{x{x'}}(t) L_{k}^x \otimes L_{k}^{x'}.
\end{eqnarray}
Combining all the results we finally arrive at
\begin{eqnarray}\label{eq:ZassenhausGenFinal}
e^{\mathcal{G}t} = e^{t(\mathcal{H}_{{\rm eff}}+\mathcal{L}_Y)} = \sum_{x,x',y,y'} d_{xx'}^{yy'} \Box _{xx'}^{yy'} + \sum_{x,x'}  \frac{\sinh\sqrt{\mathcal{A}_{xx'}(t)}}{\sqrt{\mathcal{A}_{xx'}(t)}} \sum_{k\in k_Y}A_{k}^{xx'}(t) L_{k}^x \otimes L_{k}^x
\end{eqnarray}
where the $d_{xx'}^{yy'}$ are some parameters whose explicit expressions play no role in anything that follows. Analogously it can be shown that 
\begin{eqnarray}\label{eq:ZassenhausGenFinalBack}
e^{\mathcal{\bar{G}}^{\dagger}t} = e^{t(\mathcal{H}_{{\rm eff}}+\mathcal{L}_Y^{\dagger})} = \sum_{x,x',y,y'} d_{xx'}^{yy'} \Box _{xx'}^{yy'} + \sum_{x,x'}  \frac{\sinh\sqrt{\mathcal{A}_{xx'}(t)}}{\sqrt{\mathcal{A}_{xx'}(t)}} \sum_{k\in k_Y}A_{k}^{xx'}(t) L_{k}^x \otimes L_{k}^x e^{-\Delta s_{k}},
\end{eqnarray}
the only difference being in the $e^{-\Delta s_{k}}$ factors.

So far these results are very general and apply for any choice of $\gamma_X$ and $\gamma_Y$ (as long as $\lambda = 0$). The crucial insight that allows us to make further progress is to observe that as noted in the main text, for $\gamma_X=0$ the visible jumps connect definite initial and final states allowing us to write the visible jump superoperators as $\J_k = \gamma_k^2 \kket{\chi_k^f}\bbra{\chi_k^i}$ (with $\kket{\chi_k^{i/f}}\in\{\kket{g,0},\kket{g,1},\kket{e,0},\kket{e,1}\}$). This allows us to break up Eq. (\ref{eq:DeltaIY}) for the full trajectory into $(M+1)$ individual chunks where each interval between visible transitions can be treated separately
\begin{eqnarray} 
{\Delta \sigma}_Y (\tau_X) &=& \log\frac{ \bbra{b}  e^{\mathcal{G}\Delta t_{M}} \mathcal{J}_{k_{M}}...\mathcal{J}_{k_1} e^{\mathcal{G}\Delta t_0}\kket{a}}{\bbra{b}  e^{\mathcal{\bar{G}^{\dagger}}\Delta t_{M}} \mathcal{J}_{k_{M}}...\mathcal{J}_{k_1} e^{\mathcal{\bar{G}^{\dagger}}\Delta t_0}\kket{a}} \\
&=& \log \prod_{n=0}^M \frac{\bbra{\chi_{k_{n+1}}^i}  e^{\mathcal{G}\Delta t_{n}} \kket{\chi_{k_{n}}^f}}{\bbra{\chi_{k_{n+1}}^i}  e^{\mathcal{\bar{G}}^{\dagger}\Delta t_{n}} \kket{\chi_{k_{n}}^f}}\\
\label{eq:DeltaIYSplit} &=& \sum_{n=0}^M {\Delta \sigma}_Y^n(\tau_X)
\end{eqnarray} 
where we have defined the hidden entropy production of the $n$th interval
\begin{eqnarray}\label{eq:DeltaIYn}
{\Delta \sigma}_Y^n(\tau_X) := \log \frac{\bbra{\chi_{k_{n+1}}^i}  e^{\mathcal{G}\Delta t_{n}} \kket{\chi_{k_{n}}^f}}{\bbra{\chi_{k_{n+1}}^i}  e^{\mathcal{\bar{G}}^{\dagger}\Delta t_{n}} \kket{\chi_{k_{n}}^f}}.
\end{eqnarray}
We can now substitute Eqs. (\ref{eq:ZassenhausGenFinal}) and (\ref{eq:ZassenhausGenFinalBack}) and study analyse the expression for ${\Delta \sigma}_Y^n(\tau_X)$. Let us first consider the case where $\kket{\chi^i_{k_{n+1}}} = \kket{\chi^f_{k_{n}}}$. This means that the $(n+1)$th visible transition starts in the same state that the $n$th transition ended. Intuitively this tells us that in this case ${\Delta \sigma}_Y^n$ should be zero, since whatever happened in the hidden system, it was completely balanced, equivalent to nothing happening at all. This can easily be verified by looking at Eqs. (\ref{eq:ZassenhausGenFinalBack}) and (\ref{eq:ZassenhausGenFinal}). If $\kket{\chi^i_{k_{n+1}}} = \kket{\chi^f_{k_{n}}}$, the only relevant term in $e^{\mathcal{G}t}$ and $e^{\mathcal{\tilde{G}}^{\dagger}t}$ is one of the diagonal $\Box _{xx'}^{yy'}$ terms, which has the same $d_{xx'}^{yy'}$ coefficient in both expressions, thus resulting in ${\Delta \sigma}_Y^n = \ln\frac{d_{xx'}^{yy'}}{d_{xx'}^{yy'}} = 0$ as expected.
If on the other hand $\kket{\chi^i_{k_{n+1}}} \neq \kket{\chi^f_{k_{n}}}$ we know that some hidden transitions must have taken place and thus in general expect ${\Delta \sigma}_Y^n \neq 0$. We can immediately ignore all the $\Box _{xx'}^{yy'}$ terms in Eqs. (\ref{eq:ZassenhausGenFinalBack}) and (\ref{eq:ZassenhausGenFinal}) since they only connect equal states. Further, since $\kket{\chi^f_{k_{n}}}$ has a known system state of either $\ket{e}$ or $\ket{g}$, the sums over $x$ and $x'$ reduce to a single term since only the $x=x'=e$ or $x=x'=g$ term is non-vanishing. Finally, since $\kket{\chi^f_{k_{n}}}$ and $\kket{\chi^i_{k_{n+1}}}$ also have a known demon state of either $\ket{0}$ or $\ket{1}$, the sum over $k_Y$ reduces to only the terms whose associated operators induce the correct transition in the demon states, either $\ket{0}\rightarrow\ket{1}$ (in our case $L_6$ and $L_8$) or $\ket{1}\rightarrow\ket{0}$ (in our case $L_5$ and $L_7$). As noted above, these terms that are associated with the same state to state transition (i.e. here the pairs 5/7 and 6/8) also have the same $A_{k}^{xx'}(t)$. Hence the $A_{k}^{xx'}(t)$ also cancel in the ratio in ${\Delta \sigma}_Y^n$. What we are left with is the expression
\begin{eqnarray}\label{eq:DeltaIReduced}
{\Delta \sigma}_Y^n(\tau_X) &=& \log \frac{\bbra{\chi_{k_{n+1}}^i}  \sum_{k\in k_Y} L_{k} \otimes L_{k} \kket{\chi_{k_{n}}^f}}{\bbra{\chi_{k_{n+1}}^i}  \sum_{k\in k_Y} L_{k} \otimes L_{k} e^{-\Delta s_{k}} \kket{\chi_{k_{n}}^f}} \\
&=& \log \frac{ \sum_{k\in k_Y}|\bra{\chi_{k_{n+1}}^i}   L_{k} \ket{\chi_{k_{n}}^f}|^2}{\sum_{k\in k_Y}e^{-\Delta s_{k}} |\bra{\chi_{k_{n+1}}^i}   L_{k} \ket{\chi_{k_{n}}^f}|^2 } \\
&=& - \log \frac{\sum_{k\in k_Y}e^{-\Delta s_{k}} |\bra{\chi_{k_{n+1}}^i}   L_{k} \ket{\chi_{k_{n}}^f}|^2 }{\sum_{k\in k_Y}|\bra{\chi_{k_{n+1}}^i}   L_{k} \ket{\chi_{k_{n}}^f}|^2} \\
&=& - \log \sum_{k\in k_Y} \frac{|\bra{\chi_{k_{n+1}}^i}   L_{k} \ket{\chi_{k_{n}}^f}|^2 }{\sum_{k'\in k_Y}|\bra{\chi_{k_{n+1}}^i}   L_{k'} \ket{\chi_{k_{n}}^f}|^2} e^{-\Delta s_{k}} \\
&=& - \log \sum_{k\in k_Y} p_k^n(\tau_X) e^{-\Delta s_{k}}.
\end{eqnarray}
Substituting this into (\ref{eq:DeltaIYSplit}) we arrive at (\ref{DeltaID_medium}). This concludes the proof. As noted in the main text, the result (\ref{DeltaID_simple}) is just a special case of (\ref{DeltaID_medium}) and also follows directly from the above considerations. 

\end{document}